\documentclass[preprint,aps,nofootinbib,UTF8]{revtex4}
\usepackage{graphicx}
\pdfoutput=1
\usepackage{epsfig}
\usepackage{amsmath}
\usepackage{amsfonts}
\usepackage{amssymb}
\usepackage{color}
\usepackage{dcolumn}
\usepackage{multirow}
\usepackage{booktabs}
\usepackage{subfigure}
\usepackage{CJKutf8}
\usepackage{mathrsfs}
\usepackage{graphics}
\usepackage{float}
\usepackage{indentfirst}
\usepackage{hyperref}
\providecommand{\U}[1]{\protect\rule{.1in}{.1in}}
\newcolumntype{L}{>{$}l<{$}}
\newcolumntype{R}{>{$}r<{$}}

\newcommand{\f}{\begin{equation}}
\newcommand{\ff}{\end{equation}}
\newcommand{\fa}{\begin{eqnarray}}
\newcommand{\ffa}{\end{eqnarray}}

\begin{document}
\title{Quasinormal modes of regular black holes with sub-Planckian curvature and Minkowskian core}
\author{Chen Tang$^{1}$}
\email{ctangphys@stu.cwnu.edu.cn} 
\author{Yi Ling $^{2,3,1}$}
\email{lingy@ihep.ac.cn(Corresponding author)}
\author{Qing-Quan Jiang$^{1}$}
\email{qqjiangphys@yeah.net(Corresponding author)}
\author{Guo-Ping Li$^{1}$}
\email{gpliphys@yeah.net(Corresponding author)} \affiliation{$^1$ School of Physics and Astronomy, China West Normal University, Nanchong 
637002, China \\$^2$ Institute of High Energy
Physics, Chinese Academy of Sciences, Beijing 100049, China\\$^3$
School of Physics, University of Chinese Academy of Sciences,
Beijing 100049, China}

\begin{abstract}  
We investigate the perturbation of the scalar field as well as the electromagnetic field over a sort of regular black holes which are characterized by the sub-Planckian curvature and the Minkowskian core. Specifically, we compute the quasinormal modes(QNMs) by employing the pseudo-spectral method. The outburst of overtones is manifestly observed in the QNMs of these regular black holes, which can be attributed to the deviation of the Schwarzschild black hole by quantum effects of gravity. Furthermore, the QNMs under the perturbation of electromagnetic field exhibit smaller real and imaginary parts than those under scalar field perturbation. 
By comparing  the QNMs of the regular black hole featured by  Minkowskian core with those of Bardeen black hole featured by de Sitter core, we find they may be an effective tool to distinguish these BHs. 
\end{abstract}
\maketitle
\section{Introduction}
The quasinormal modes (QNMs) as the intrinsic feature of a black hole have been playing a vital role in black hole physics (For  comprehensive review on QNMs, we refer to recent Ref.\cite{Nollert:1999ji,kokkotas1999quasi,berti2009quasinormal,konoplya2011quasinormal}). By considering the perturbations surrounding a black hole and computing the QNMs, one may acquire the parameter information of a black hole such as the mass and spin\cite{echeverria1989gravitational,flanagan1998measuring,yang2012quasinormal}. Furthermore, one may diagnose the stability of a black hole by the feature of its QNMs\cite{horowitz2000quasinormal,konoplya2007stability,konoplya2009instability,Fontana:2022whx,Berti:2022xfj,Kyutoku:2022gbr,Sarkar:2023rhp,Fontana:2023dix,Arean:2023ejh,Courty:2023rxk,skvortsova2024quasinormal}. More importantly, QNMs is an important tool for analyzing and interpreting gravitational wave signals generated by the merger of binary black holes\cite{Berti:2016lat,Giesler:2019uxc,JimenezForteza:2020cve,Maggio:2020jml,Dey:2020pth,Carullo:2021oxn,Yi:2024elj,rosato2024ringdown}. 
In addition, one can use QNMs to investigate the internal structure and dynamical behavior of extremal compact objects such as neutron stars \cite{Nollert:1999ji,nollert2000characteristic,Ferrari:2007dd,Berti:2018vdi}. In particular, since the first signal of gravitational waves was successfully detected by LIGO in 2015 \cite{Abbott2016,abbott2019tests} and the first image of the black hole at the center of the M87 galaxy was captured by EHT in 2019 \cite{psaltis2020gravitational}, the QNMs have been becoming the powerful tool for linking the theory on black holes to the data of observation. The analysis on QNMs is of great significance for exploring the nature of black holes and for testing various theories on gravity by astronomical observation.  

Recently, the QNMs of regular black holes have been extensively investigated with the purpose of revealing the internal structure of black holes and capturing the possible signals due to the quantum effects of gravity \cite{berti2006gravitational,bronnikov2012instabilities,flachi2013quasinormal,rincon2020greybody,jusufi2021quasinormal,lan2021quasinormal,hendi2021physical,konoplya2023quasinormal,meng2023gravito,lopez2023quasi,gingrich2024quasinormal,simpson2021ringing,konoplya2024infinite,franchini2024testing,franzin2024regular,zhang2024quasinormal}.  Regular black holes are those  without singularity at the center of the black hole \cite{bardeen1968non,dymnikova1992vacuum,hayward2006formation,frolov2014information}, thus not suffering from the singularity problem as those black holes in standard general relativity \cite{Penrose:1964wq,ch5-}. Originally, regular black holes are proposed by purely phenomenological consideration, with the assumption that the singularity would be removed by the quantum effect of gravity\cite{bardeen1968non,dymnikova1992vacuum,hayward2006formation,frolov2014information,Xiang:2013sza,Li:2016yfd,Bianchi:2018mml,Simpson:2018tsi,Feng:2023pfq}.  Later, people find that they can also be constructed as the solutions to Einstein equations with exotic matter which locally violates the energy conditions\cite{ayon1998regular,balart2014regular,koch2014black,ayon2000bardeen,fan2016construction}, or to the effective Einstein equations which take the quantum correction of gravity into account\cite{Ashtekar:2018lag,Ashtekar:2018cay,achour2018polymer,Feng:2024sdo} (For recent review on regular black holes, we refer to Refs.\cite{ashtekar2023regular,lan2023regular}).  By perturbative analysis, it is quite interesting to notice that the QNMs of regular black holes exhibit some distinct behavior from those of black holes with singularity, thus in principle one could diagnose if the observed astrophysical black holes contain the singularity inside by examining their QNMs in future. For instance, the outburst of overtones is manifestly observed in the QNMs of regular black holes, which might be ascribed to the quantum effects of gravity \cite{berti2003asymptotic,zhu2024quasinormal,Gong:2023ghh,konoplya2022quasinormal,konoplya2023bardeen,Konoplya:2023ppx,qian2024universality,Stashko:2024wuq,Gingrich:2024tuf,konoplya2024overtones,Moreira:2023cxy,fu2024peculiar,Franchini:2022axs}. Recently, a new sort of regular black holes with an exponentially suppressing Newton potential are proposed in \cite{ling2023regular}, which is characterized  by the sub-Planckian curvature and the Minkowskian core. Sequentially, the observational feature of these regular black holes are investigated in \cite{LingLingYi:2021rfn,Ling:2022vrv,Simpson:2021biv,zeng2023accretion,zeng2023astrophysical}. Specially, the QNMs of a specific regular black hole which has the same asymptotic behavior with Hayward black hole at infinity  is investigated in \cite{zhang2024quasinormal}.
In this paper we intend to push this approach forward by considering the QNMs of a regular black hole which has the same asymptotic behavior with Bardeen black hole. The QNMs of both the scalar field and the electromagnetic field will be investigated. More importantly, we will compare the QNMs of this sort of regular black holes with different deviation parameters. It is expected that the current work will provide more 
information about the QNMs of this sort of regular black holes and then applicable to distinguish regular black holes from singular black holes by the observation in near future.  

The paper is organized as follows. In Section \ref{Regular BHs with sub-Planckian curvature}, we briefly review the regular black holes with Minkowskian core and sub-Planckian curvature originally proposed in \cite{ling2023regular}. Then in Section \ref{Quasinormal modes of regular BHs}, we investigate the QNMs of a typical regular black hole with specific parameters under scalar and electromagnetic field perturbations in detail. The QNMs of the regular black hole featured by  Minkowskian core with those of Bardeen black hole featured by de Sitter core is compared in Section \ref{Comparison between QNMs of regular BHs and Bardeen BHs}. In Section \ref{Comparison of QNMs of regular BHs with different parameters X and n under scalar field perturbations}, we compare the QNMs of regular black holes with different parameters under scalar field perturbations.  We summarize our results and discuss the possible progress in future in Section \ref{CONCLUSION AND DISCUSSION}.

\section{Regular black holes with Minkowskian core and sub-Planckian curvature}
\label{Regular BHs with sub-Planckian curvature}
\begin{figure}[ht]
	\centering{
		\includegraphics[width=8cm]{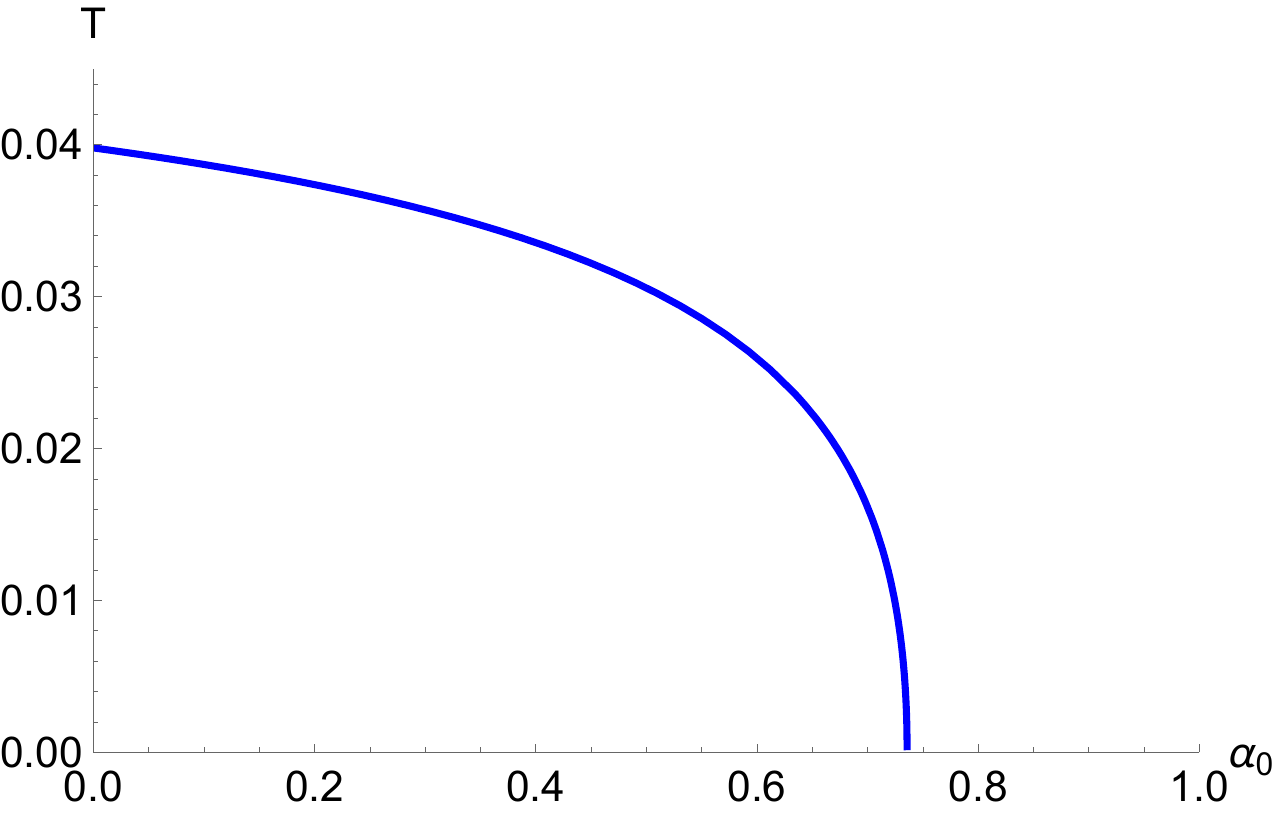}
		\caption{The Hawking temperature as a function of $\alpha_0$, where $r_h=1$ and $M=1$.}
		\label{Hawking_T}
	}
\end{figure}

In this section, we briefly review the spherically symmetric regular black hole with sub-Planckian curvature originally proposed in \cite{ling2023regular}. This sort of regular black holes is characterized by an asymptotically Minkowski core as well as an exponentially suppressing
gravity potential. Specifically, their metric  takes the following form 
\begin{equation} 
    d s^{2}=-f(r) d t^{2}+\frac{1}{f(r)}d r^{2}+r^{2}\left(d \theta^{2}+\sin ^{2} \theta d \phi^{2}\right),
    \label{metric}
\end{equation}
where the function $f(r)$ is given by
\begin{equation} 
    f(r)=1+2\psi(r)=1-\frac{2M}re^{\frac{-\alpha_0M^x}{r^c}},
    \label{psi}
\end{equation}
where $M$ is the mass of black hole and the parameters $x$, $c$, $\alpha_0$ are all the dimensionless parameters \footnote{We have set $8\pi G=l_p^2=1$ throughout this paper for convenience. It means if one intends to recover the dimension of the potential, it should become as $\psi(r)=-\frac{MG}re^{\frac{-\alpha_0(MG)^xl_p^{c-x}}{r^c}}$.}. Importantly, $\alpha_0$ indicates the degree of deviation from the Newtonian potential and characterizes the corrections due to quantum gravity effects\cite{Xiang:2013sza,Li:2016yfd}. Obviously, when $\alpha_0=0$, the potential reduces to the standard Newtonian potential and this regular BH goes back to an ordinary Schwarzschild BH. In addition, to preserve the spacetime curvature to be sub-Planckian, the parameters should satisfy the conditions $c \geq x \geq c/3$ and $c\geq 2$ \cite{ling2023regular}. Previously, the QNMs of the scalar field over regular black hole with $x=1$ and $c=3$ was studied in \cite{zhang2024quasinormal}, with a comparison with the QNMs of Hayward black hole. 
In this paper, we will extend the above analysis to the QNMs of both the scalar field and the electromagnetic field over the regular black hole with $x=2/3$ and $c=2$ at first, which has the same asymptotic behavior at infinity as the Bardeen BH. Moreover, we will compare the QNMs of this sort of regular black holes with general parameters $x$ and $c$ under the above sub-Planckian conditions.    
For the metric with the function in (\ref{psi}), the event horizon can be obtained by solving the equation $f(r_h)=0$, and $r_h$ is the location of the event horizon. Then, it is straightforward to derive the Hawking temperature as
\begin{equation} 
T=\frac{f'(r_h)}{4\pi}=\frac{\psi'(r_h)}{2\pi}.
\label{Tem1}
\end{equation}
By substituting Eq.(\ref{psi}) into Eq.(\ref{Tem1}), the specific form of Hawking temperature can be expressed as
\begin{equation} 
    T=\frac{e^{-\frac{M^{2/3}\alpha_0}{r_h^2}}M\left(r_h^2-2M^{2/3}\alpha_0\right)}{2\pi{r_h}^4}. 
    \label{Tem2}
\end{equation}
Here, the Hawking temperature should be greater than zero, implying that the deviation parameter $\alpha_0$ is constrained as
\begin{equation} 
	\alpha_0\leq\frac{2M^\frac{4}{3}}{e}.
 \label{alpha}
\end{equation}
In FIG. \ref{Hawking_T}, we plot the Hawking temperature as a function of the deviation parameter $\alpha_0$ with $M=1$. It can be seen that the temperature $T$ decreases as $\alpha_0$ increases until it drops to zero at $\alpha_0=2/e$. 
Finally, we remark that the regular black hole with $x=2/3$ and $c=2$  has the same asymptotic behavior at infinity as the Bardeen black hole. At large scales where $r\gg\sqrt{\alpha_{0}}M^{1/3}$, the function $f(r)$ can be expanded as the following form
\begin{equation}
f(r)=1+2\psi(r)=1-\frac{2M}{r}e^{-\alpha_0M^{2/3}/r^2}\cong1-\frac{2M}{r}(1-\frac{\alpha_0M^{2/3}}{r^2}+...).
\end{equation}
On the other hand, the function $\psi$ of the Bardeen black hole has the form
\begin{equation}
\psi(r)=-\frac{Mr^2}{\left(\frac{2}{3}\alpha_0M^{2/3}+r^2\right)^{3/2}}.
\end{equation}
So, at large scales the function $f(r)$ asymptotically behaves as 
\begin{equation}
f(r)=1+2\psi(r)=1-\frac{2Mr^2}{\left(\frac{2}{3}\alpha_0M^{2/3}+r^2\right)^{3/2}}\cong1-\frac{2M}{r}(1-\frac{\alpha_0M^{2/3}}{r^2}+...),
\end{equation}
which is the same as the regular black hole with $x=2/3$ and $c=2$. Nevertheless, we point out that these two black holes  exhibit distinct behavior at the center of the black hole. As $r\rightarrow 0$, the regular black hole with $x=2/3$ and $c=2$ has a Minkowskian core while Bardeen black hole has a de-Sitter core. In this paper we are concerned with the behavior of QNMs in these two regular black holes when the parameter $\alpha_0$ takes the same value. 

\section{Quasinormal modes of the regular black hole with $x=2/3$ and $c=2$ }
\label{Quasinormal modes of regular BHs}
In this section, we study the QNMs of the regular black hole with $x=2/3$ and $c=2$ under the perturbation of the scalar and electromagnetic field, respectively. In general,
the asymptotic behavior of the QNM close to the boundary should satisfy
\begin{subequations}
\begin{align}
   \Psi &\sim e^{-i\omega(t+r_*)}\! & r_* &\to +\infty, \\
     \Psi &\sim e^{-i\omega(t-r_*)}\! & r_* &\to -\infty,
\end{align}
\end{subequations}
where  $r_*$ is the ordinary radial tortoise coordinate. This indicates that the QNMs at the event horizon are purely ingoing waves, while at infinity are purely outgoing waves. In this section, we consider the perturbation of  scalar field and electromagnetic field, which are described by the Klein-Gordon equation  and the Maxwell equations, respectively
\begin{subequations}
\label{eq:main}
\begin{align}
    \frac1{\sqrt{-g}}(g^{\mu\nu}\sqrt{-g}\partial_\mu\Phi),_\nu &= 0,  \label{klein_alpha} \\
    \frac1{\sqrt{-g}}(g^{\alpha\mu}g^{\sigma\nu}\sqrt{-g}F_{\alpha\sigma}),_\nu &= 0  \label{maxwell_alpha},
\end{align}
\end{subequations}
where $g^{\mu\nu}$ is the contravariant form of the metric, and $g$ is its determinant. By virtue of the spherical symmetry of the background, 
one can separate variables and decompose  the wave function into the form of spherical harmonic functions. For instance, for the scalar field, it can be decomposed as
\begin{equation}\label{separate}
\Phi(t,r,\theta,\phi)=\sum_{l,m}Y_{l,m}(\theta,\phi)\frac{\Psi_{l,m}(t,r)}{r},
\end{equation}
where $Y_{l,m}(\theta,\phi)$ denotes the spherical harmonic function, with the symbols $l$ and $m$ representing the angular quantum number and magnetic quantum number, respectively. The similar decomposition can be performed for the Maxwell field and we refer to \cite{fu2024peculiar} for details. Substituting Eq.(\ref{separate}) into the perturbation Eq.(\ref{eq:main}), and further performing the tortoise coordinate transformation ($dr_*/dr = 1/f(r)$), one changes both the Klein-Gordon equation and Maxwell equation into a Schrödinger-like form, which is 
\begin{equation}\label{Sch_like_eq}
   \frac{\partial^2\Psi}{\partial r_*^2}+(\omega^2-V_{eff})\Psi=0\:.
\end{equation}
We are concerned with the solution of the wave function as the values of $l$ and $m$ are given for the spherical harmonics, thus for convenience, the lower indices $l$ and $m$ are omitted in the notation of the wave function in the following. The effective potential in the Schrödinger-like equation is
\begin{equation}
V_s(r)=f(r)\left(\frac{l(l+1)}{r^2}+(1-s)\frac{f'(r)}{r}\right).
\end{equation}
When the spin $s = 0$, it corresponds to scalar perturbations; while for $s = 1$, it corresponds to electromagnetic perturbations.\par

A lot of methods have been presented in literature for one to compute the QNMs, including the Wentzel–Kramers–Brillouin (WKB) method, the pseudo-spectral(ps) method, the continued fraction method \cite{leaver1985analytic}, Horowitz-Hubeny method \cite{horowitz2000quasinormal}, asymptotic iteration method \cite{cho2010black} and the matrix method \cite{lin2019matrix,lin2017matrix}. 
In this paper we will employ pseudo-spectral method to numerically investigate the QNMs of regular black holes
\footnote{In fact, we have also employed the WKB method to study the the QNMs of regular black holes (\ref{metric}), and obtained the similar conclusion with \cite{zhang2024quasinormal}, i.e., the WKB method is not quite suitable for computing highly damped QNMs. This part of the content is not the focus of our article.}, while the details of the pseudo-spectral method can be referred to Appendix \ref{sec:pseudospectral-method}.

\subsection{The QNMs under the perturbation of a scalar field}
Obviously, the specific form of the effective potential plays a key role in determining the QNMs of the perturbations. In FIG. \ref{veff l=0，1，2,3}, we plot the effective potential $V_{\text{eff}}(r)$ of the scalar field with different deviation parameters $\alpha_0$. 
It shows that the effective potential is always positive, indicating that the spacetime of this regular BH is stable under the scalar perturbations. Furthermore, the effective potential increases with the increase of $l$. Also, it is interesting to notice that the maximal value of the potential $V_{\text{eff}}$ decreases as $\alpha_0$ increases for $l=0$; while for $l=1,2,3$, its maximal value increases with $\alpha_0$. We remark that this special behavior of the effective potential at $l=0$ might be responsible for  the spiral behavior of QNMs that appears in the first few overtones, as illustrated in the next context.
\begin{figure}
	\centering{
		\includegraphics[width=7.2cm]{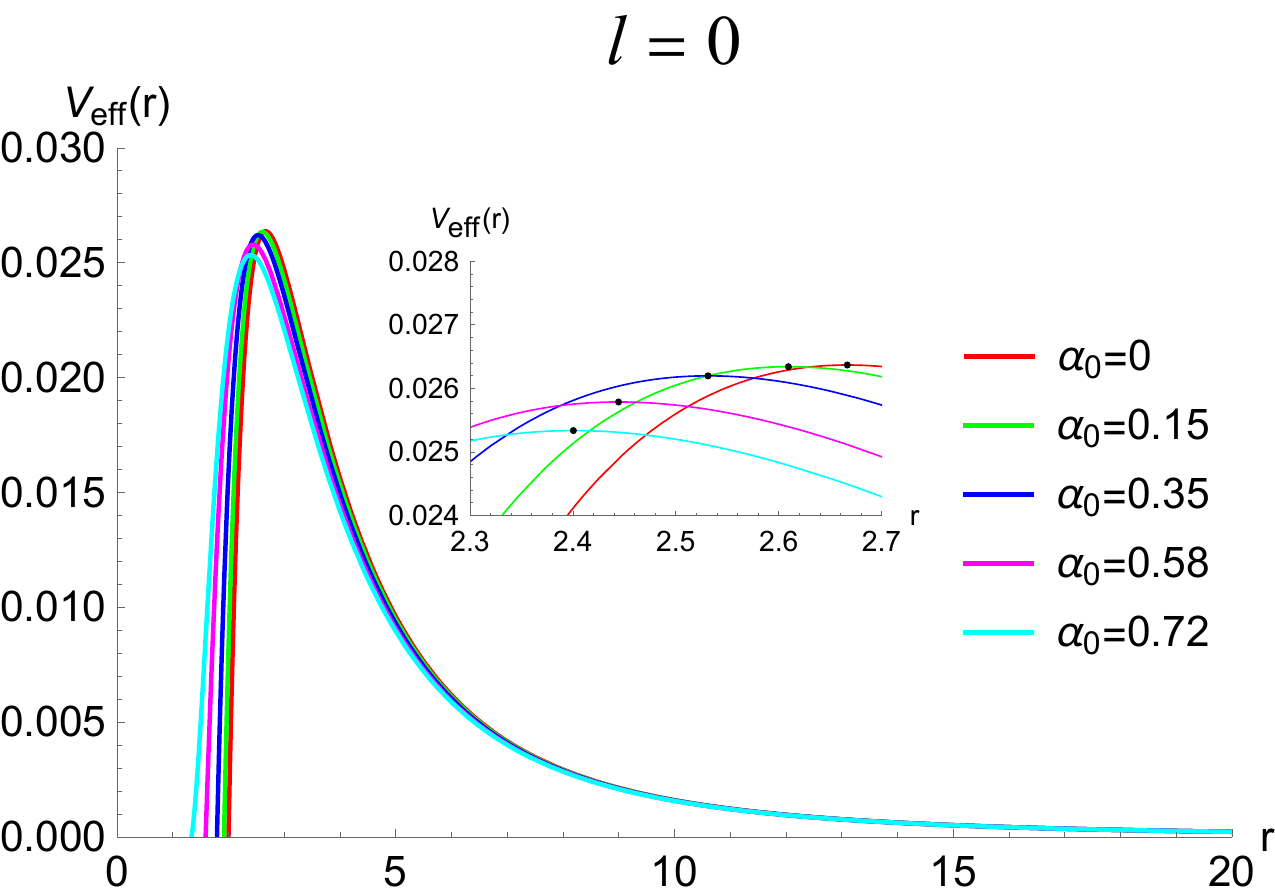}\hspace{0.5cm}
		\includegraphics[width=7.2cm]{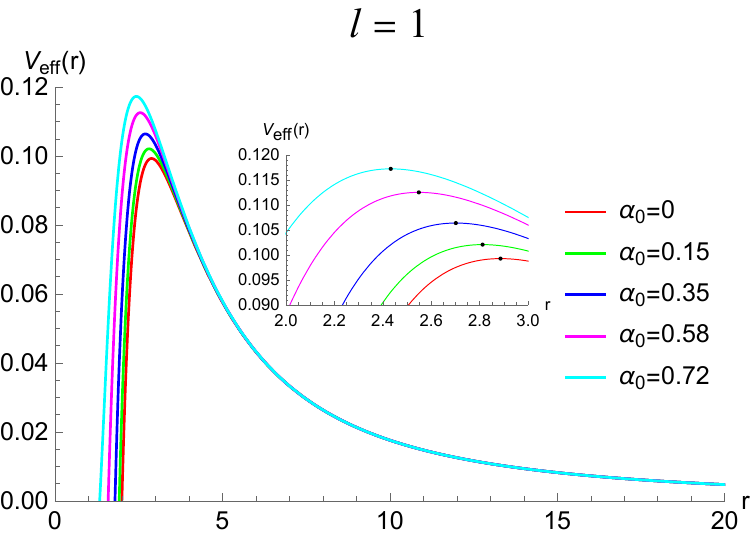}\
		\includegraphics[width=7.2cm]{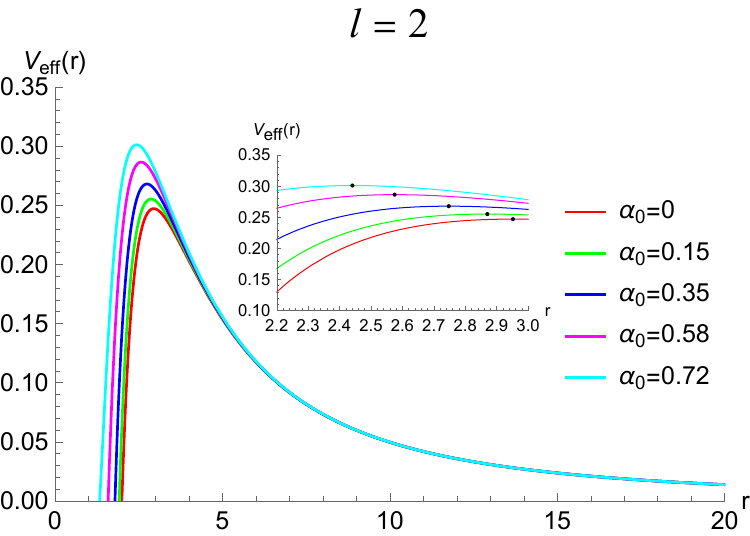}\hspace{0.5cm}
		\includegraphics[width=7.2cm]{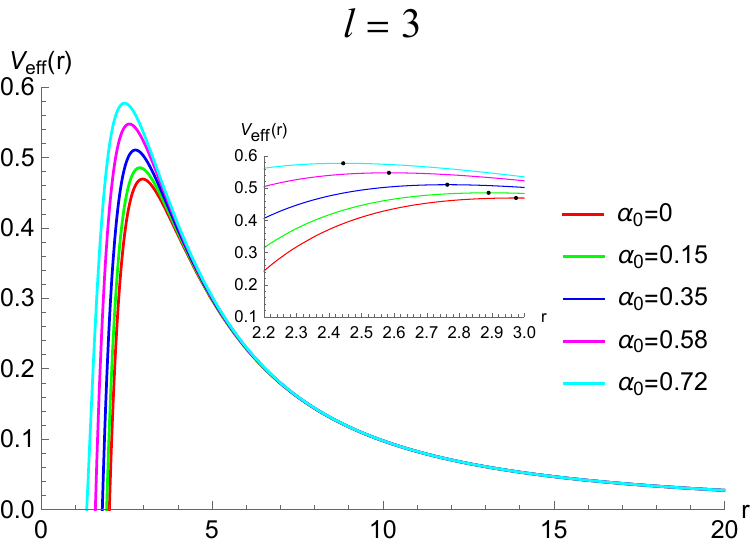}\
		\caption{The effective potential $V_{\text{eff}}(r)$ of the scalar field with different deviation parameters $\alpha_0$. The black dot represents the maximum value of the $V_{\text{eff}}(r)$.}
		\label{veff l=0，1，2,3}
	}
\end{figure}

Now we present the numerical results for the QNMs of regular black hole under the perturbation of scalar field, which are obtained with the pseudo-spectral method. In Tables (\ref{tab1 scalar l=0}), (\ref{tab2 scalar l=1}) and (\ref{tab3 scalar l=2}), we list the frequency of  the QNMs for different overtone numbers $n$ and deviation parameters $\alpha_0$. 
\begin{table}[ht]
\begin{tabular}{c@{\hspace{1em}}c@{\hspace{2em}}c@{\hspace{2em}}c@{\hspace{2em}}c} 
\hline
\hline
n & $\alpha_0=0$ & $\alpha_0=0.3$ & $\alpha_0=0.5$ & $\alpha_0=2/e$  \\ 
\hline
\hline
0 & 0.110455-0.104896i & 0.115834-0.100752i & 0.119115-0.095696i & 0.116896-0.088312i \\
1 & 0.086117-0.348053i & 0.093067-0.330182i & 0.090531-0.309631i & 0.074856-0.309121i \\
2 & 0.075776-0.601067i&0.080399-0.569011i&0.052095-0.537955i&0.055515-0.553466i\\
\hline
\hline
\end{tabular}
\caption{The frequency of QNMs under scalar perturbations for $l = 0$}.
\label{tab1 scalar l=0}
\end{table}
\begin{table}[ht]
\centering
\begin{tabular}{c@{\hspace{1em}}c@{\hspace{2em}}c@{\hspace{2em}}c@{\hspace{2em}}c} 
\hline
\hline
n & $\alpha_0=0$ & $\alpha_0=0.3$ & $\alpha_0=0.5$ & $\alpha_0=2/e$  \\ 
\hline
\hline
0 &0.292936-0.09766i & 0.304502-0.094405i &0.313652-0.090570i&0.325102-0.082546i\\
1 &0.264449-0.306257i& 0.280096-0.293848i &0.290682-0.279744i&0.294583-0.255512i\\
2 &0.229539-0.540133i& 0.248559-0.513917i &0.256659-0.485222i&0.242126-0.453212i\\
3 &0.203258-0.788298i& 0.222508-0.746671i &0.219932-0.703117i&0.190160-0.681290i\\
4 &0.185109-1.040760i & 0.202301-0.983759i &0.177099-0.930025i&0.155153-0.931619i\\
5 &0.172080-1.294130i  & 0.185453-1.221950i  &0.136329-1.195380i &0.128669-1.176260i\\
\hline
\hline
\end{tabular}
\caption{The frequency of QNMs under scalar perturbations for $l = 1$.}
\label{tab2 scalar l=1}
\end{table}
\begin{table}[ht]
\centering
\begin{tabular}{c@{\hspace{1em}}c@{\hspace{2em}}c@{\hspace{2em}}c@{\hspace{2em}}c} 
\hline
\hline
n & $\alpha_0=0$ & $\alpha_0=0.3$ & $\alpha_0=0.5$ & $\alpha_0=2/e$  \\ 
\hline
\hline
0 &0.483644-0.096759i& 0.502162-0.093662i &0.517272-0.090012i & 0.538533-0.081991i  \\
1 &0.463851-0.295604i& 0.485334-0.285272i &0.501871-0.273277i & 0.519576-0.248721i \\
2 &0.430544-0.508558i& 0.456643-0.488169i &0.474525-0.465111i &0.482746-0.424192i\\
3 &0.393863-0.738097i& 0.423946-0.704539i &0.440578-0.667589i &0.431833-0.615047i\\
4 &0.361299-0.979922i& 0.393356-0.931434i &0.404192-0.879340i&0.375024-0.825865i\\
5 &0.334900-1.228410i & 0.366972-1.164490i &  0.366463-1.098090i& 0.322525-1.054220i\\
\hline
\hline
\end{tabular}
\caption{The frequency of QNMs under scalar perturbations for $l = 2$.}
\label{tab3 scalar l=2}
\end{table}

In each table the angular momentum $l$ is fixed as $l=0$, $l=1$ and $l=2$, respectively. Firstly, it shows that the imaginary part of all QNM frequencies is always negative, indicating that the regular BH is stable under the scalar perturbations. Secondly,
we are concerned with the effects of the overtone number $n$ and quantum number $l$ on the QNMs. In each table with fixed $l$, we notice that with the increase of the overtone number $n$, the real part of QNMs decreases and the imaginary part increases, indicating that the QNMs exhibit weaker oscillations and faster decay with $n$, which
is similar to that of ordinary black holes with singularity. On the other hand, if we fix $n$ and $\alpha_0$ but change $l$, then we notice that the real part of the QNM frequency becomes larger as $l$ increases. The reason is that as $l$ increases, the effective potential becomes larger as well, as illustrated in FIG. \ref{veff l=0，1，2,3}. Thus it requires that the vibration frequency of the wave becomes larger such that the QNMs can escape from the potential and transmit to infinity. However, 
the imaginary part of the frequency does not change much as $l$ increases, implying that the attenuation rate of the QNM is majorly determined by the parameter $n$ rather than $l$.
Next, we focus on the effect of the deviation parameter $\alpha_0$, which is the major part that we are concerned with. For $l=0$ in Table (\ref{tab1 scalar l=0}), we notice that the real part increases initially and then decreases with the increase of the parameter $\alpha_0$, while for $l=1$ and $l=2$,  it becomes monotonously increasing with $\alpha_0$ for smaller $n$, but again exhibits the similar behavior as $l=0$ for larger $n$. Similarly, we find that the monotonous behavior of the imaginary part stops with the increase of the overtone number $n$. 

To reveal the behavior of the QNMs with the variation of $\alpha_0$ more transparently, we intend to plot the trajectory of the QNMs on the plane of the  complex frequency.
For each specified $n$ and $l$, we compute the frequency of the QNMs as the deviation parameter $\alpha_0$ runs from zero to $2/e$, and then plot a trajectory on the plane of the complex frequency, as illustrated in  FIG. \ref{l=0,n=1,2} to FIG. \ref{l=2,n=0，1，2，3,4,5}, where $l$ and $n$ are specified as various values. 
\begin{figure}[ht]
	\centering{
		\includegraphics[width=7.8cm]{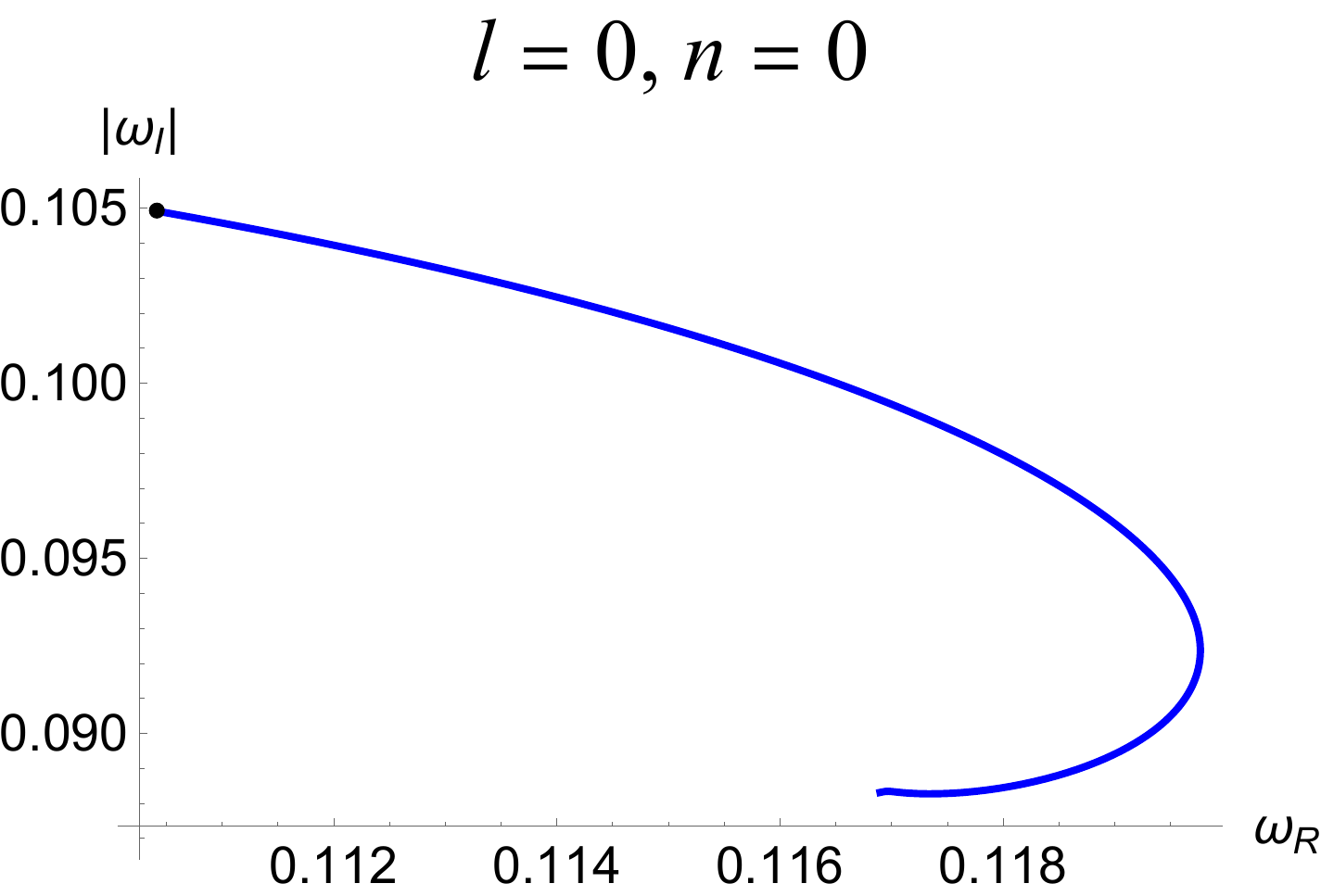}\hspace{6mm} 
		\includegraphics[width=7.8cm]{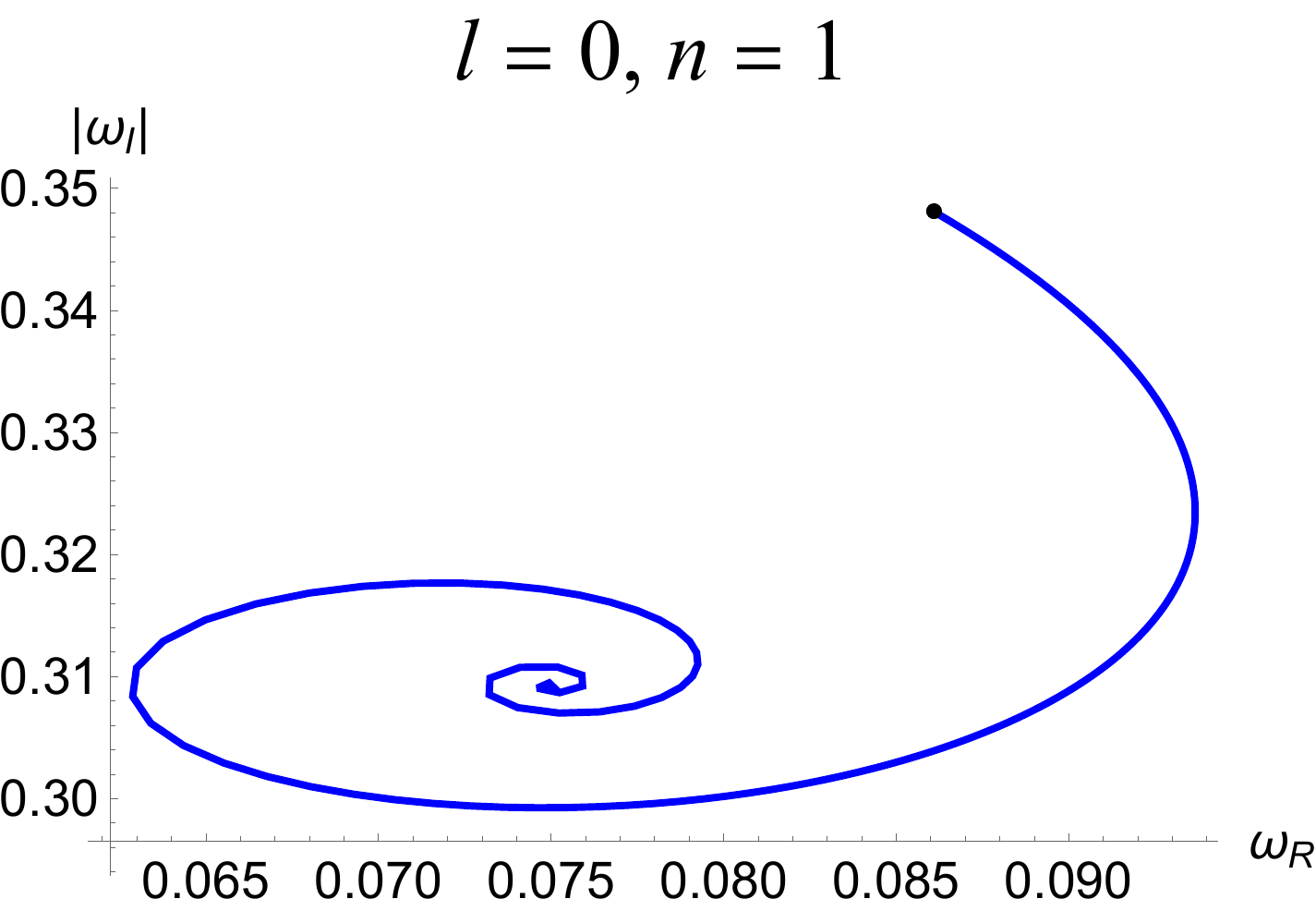}
		\caption{The trajectory of the QNMs with the variation of $\alpha_0$ on the frequency plane with $l=0$, where  $\omega_R$ is the real part and $|\omega_I|$ is the magnitude of the imaginary part. The black dots denote the QNM of Schwarzschild black hole, which is same in the following figures.
		}
		\label{l=0,n=1,2}
	}
\end{figure}

\begin{figure}[ht]
    \centering
    \includegraphics[width=5.2cm]{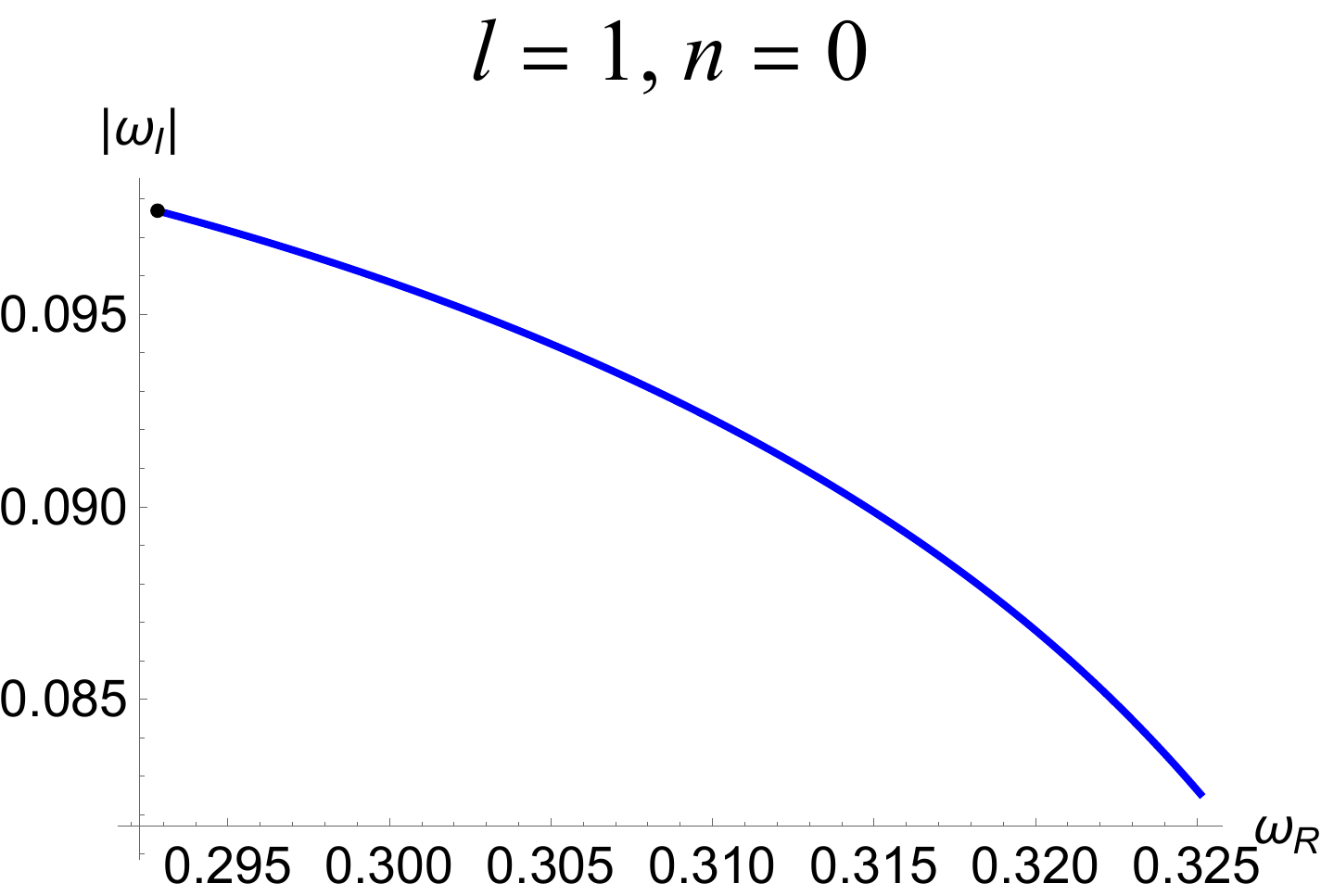}\hspace{1mm}
    \includegraphics[width=5.2cm]{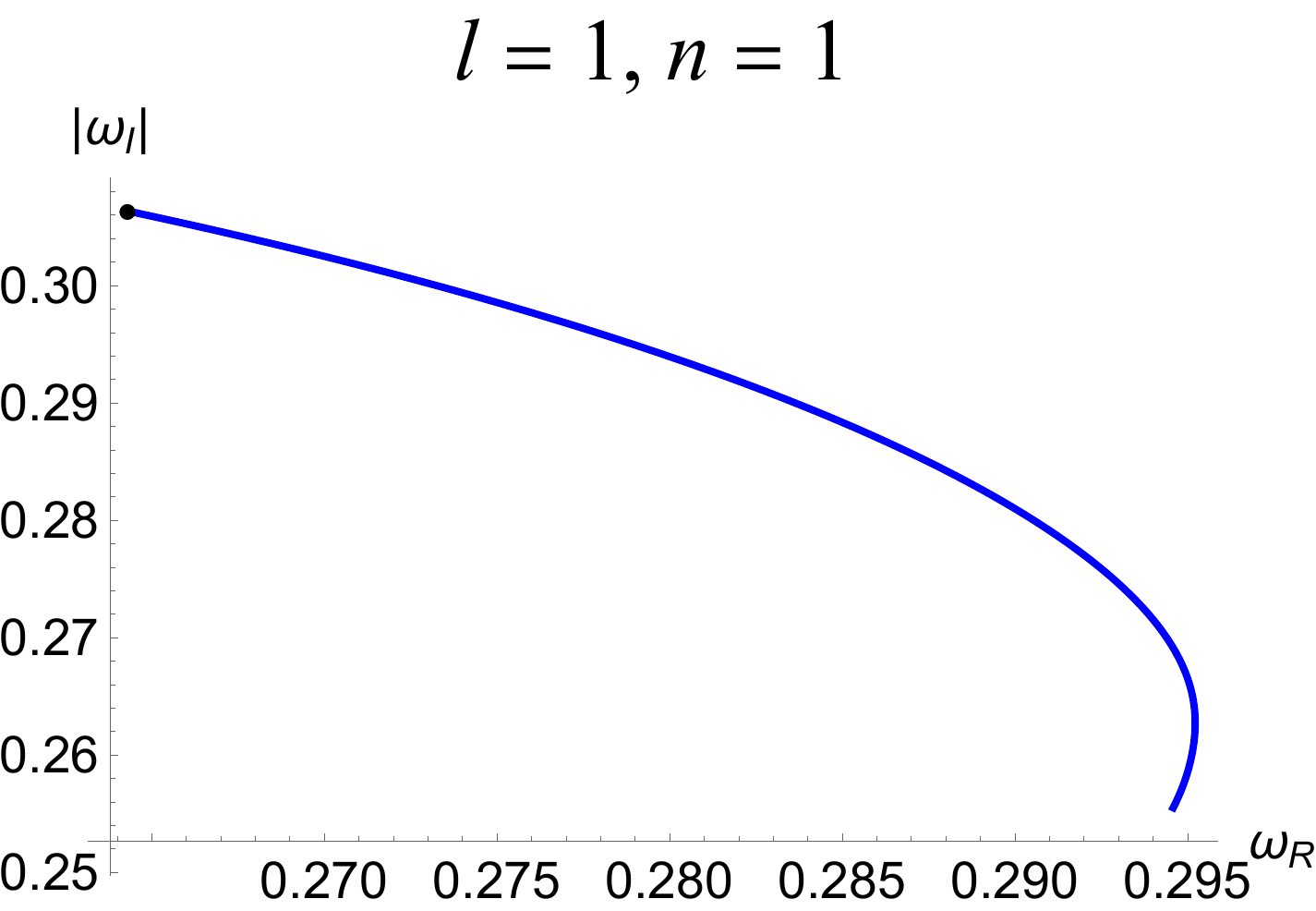}\hspace{1mm}
    \includegraphics[width=5.2cm]{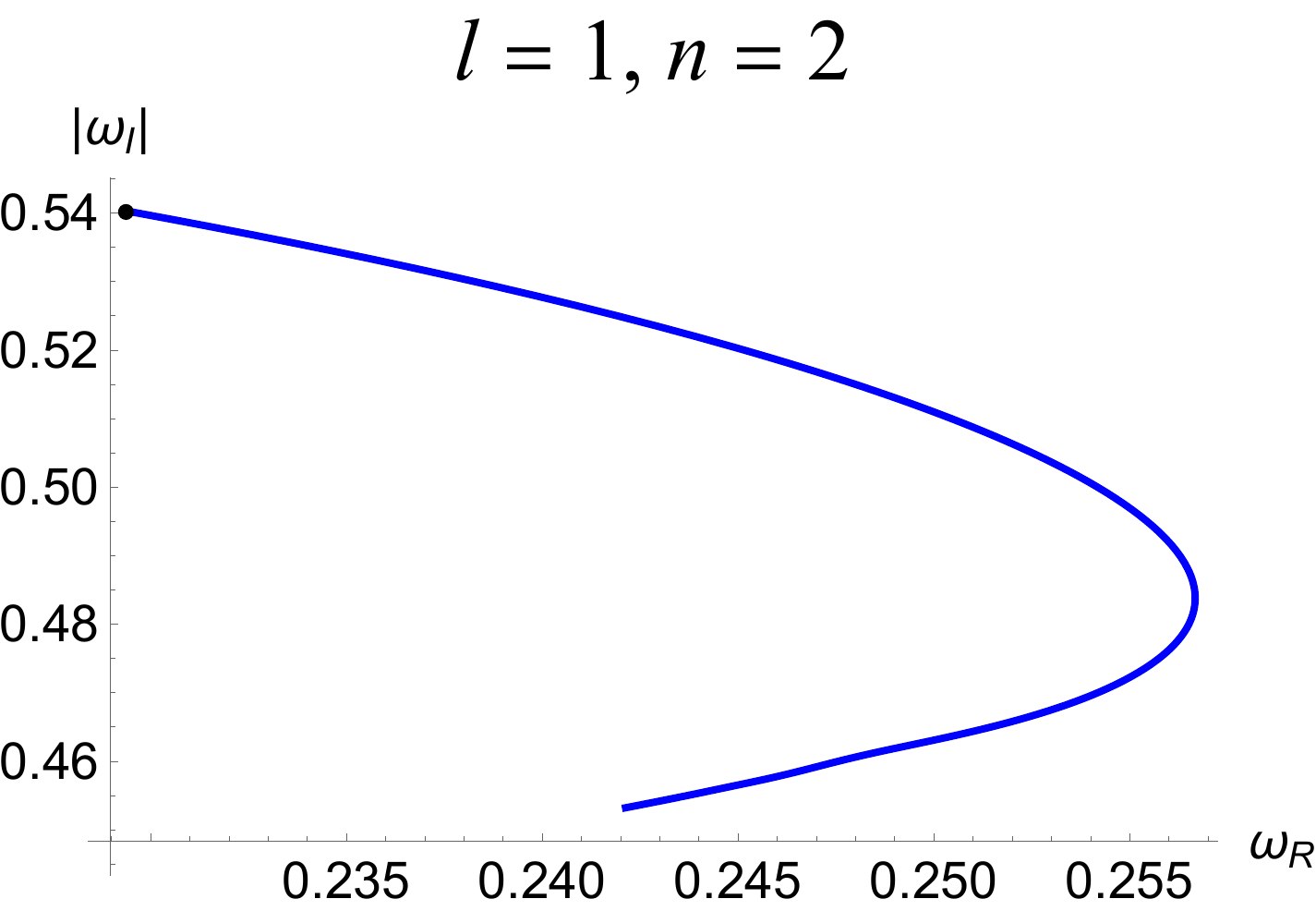}\hspace{1mm}
    \includegraphics[width=5.2cm]{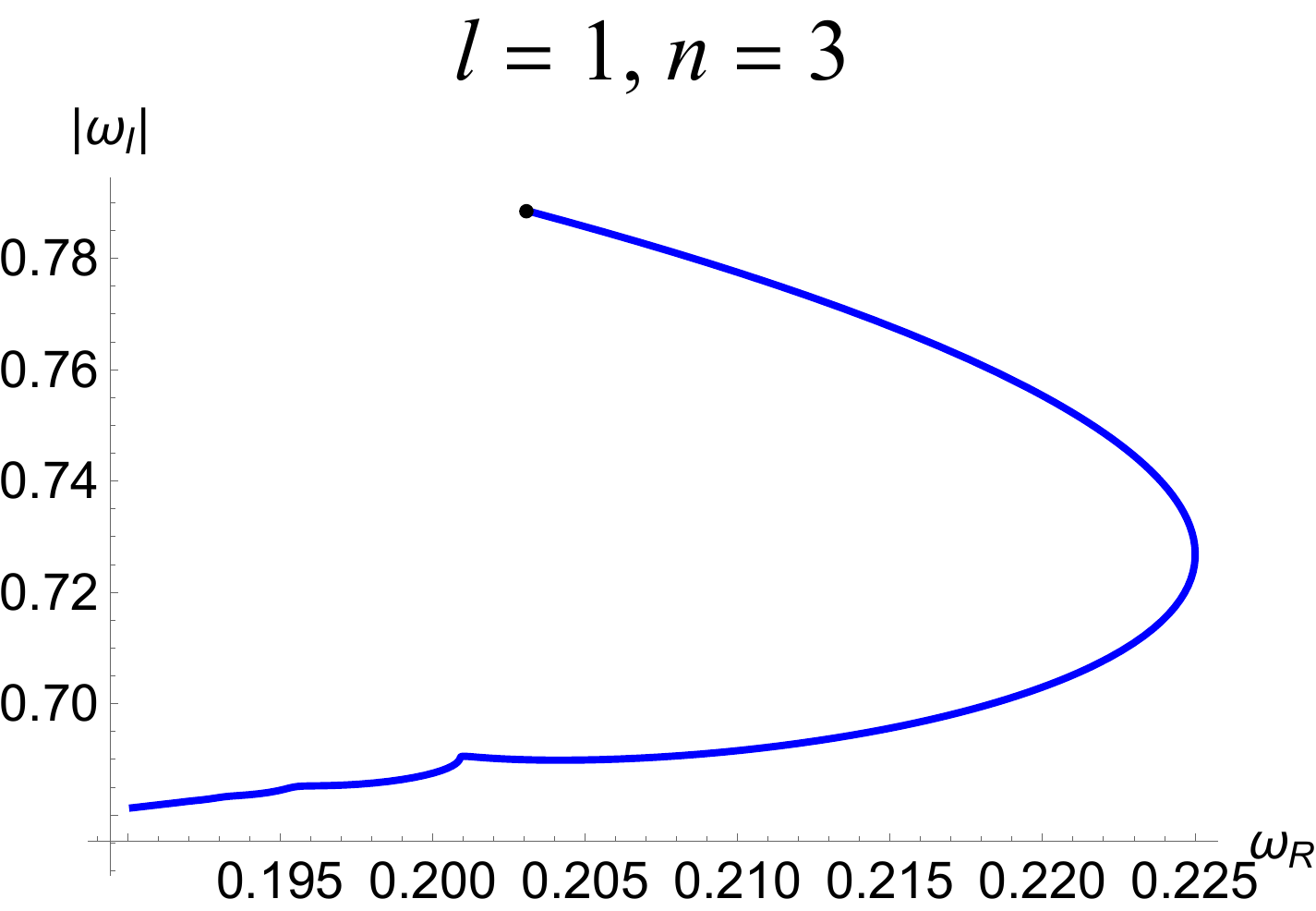}\hspace{1mm}
    \includegraphics[width=5.2cm]{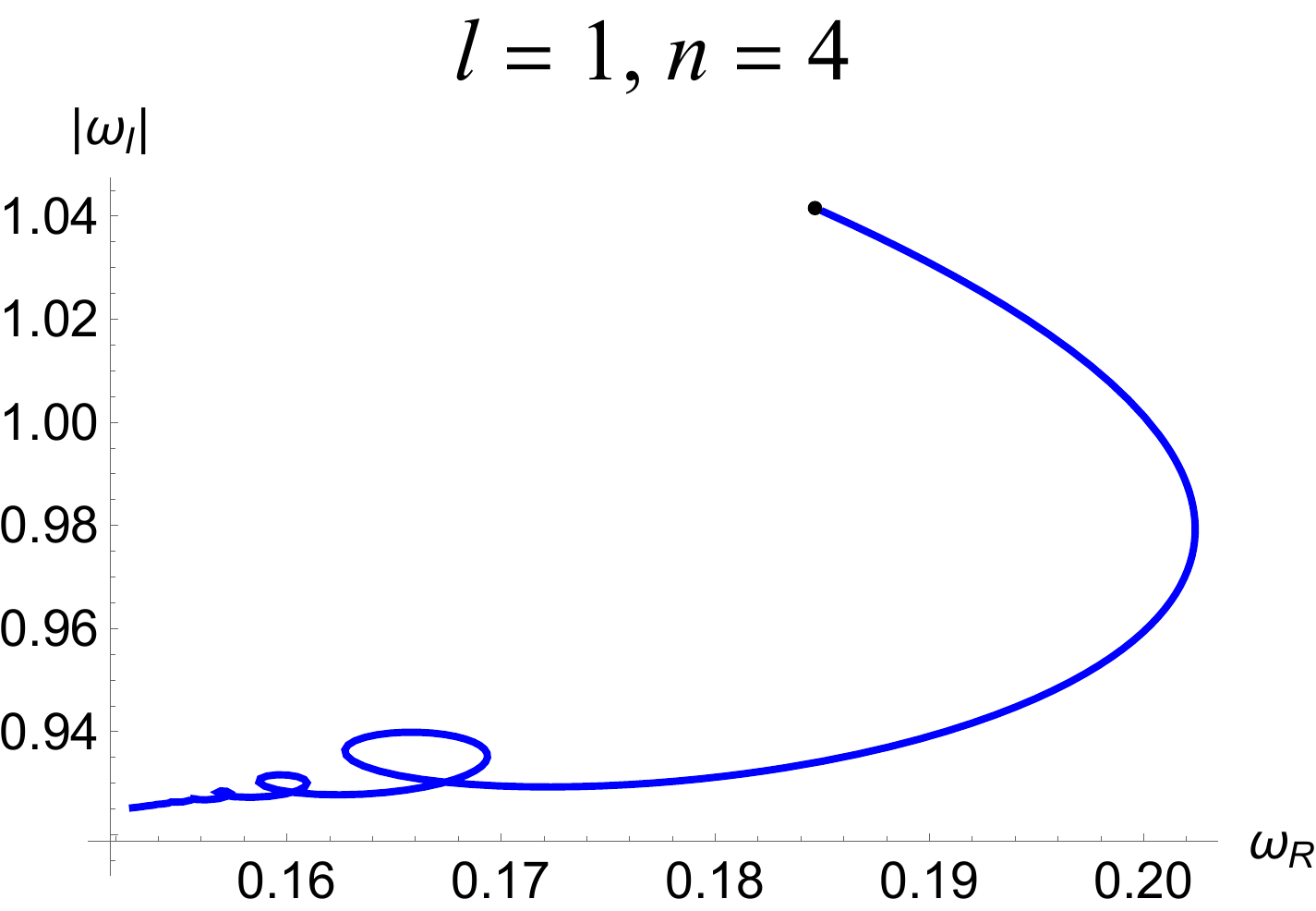}\hspace{1mm}
    \includegraphics[width=5.2cm]{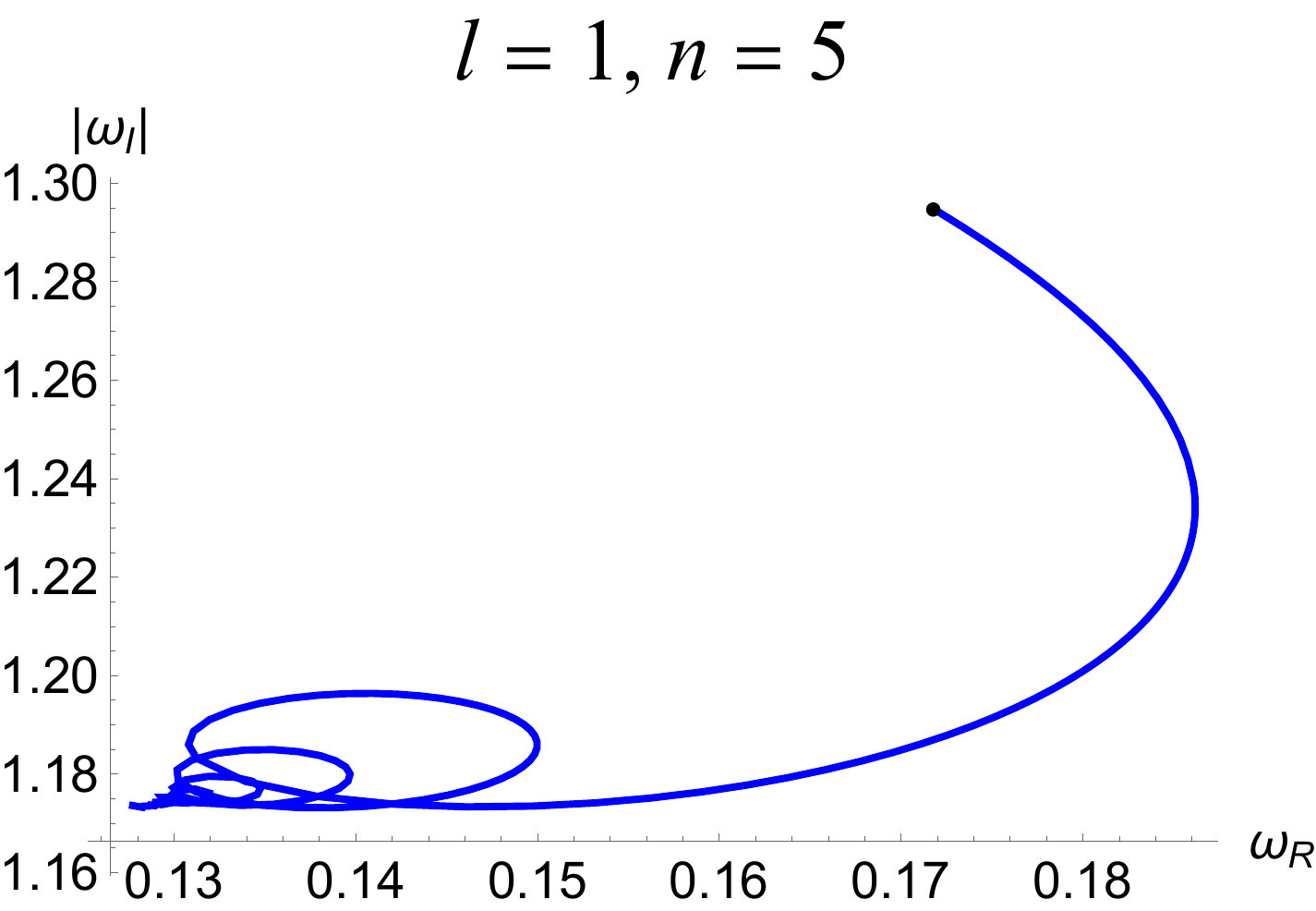}\hspace{1mm}
    \caption{The trajectory of the QNMs with the variation of $\alpha_0$ on the frequency plane with $l=1$} under scalar field perturbation.
		\label{l=1,n=0，1，2，3,4,5}
\end{figure}
\begin{figure}[ht]
	\centering{
		\includegraphics[width=5.2cm]{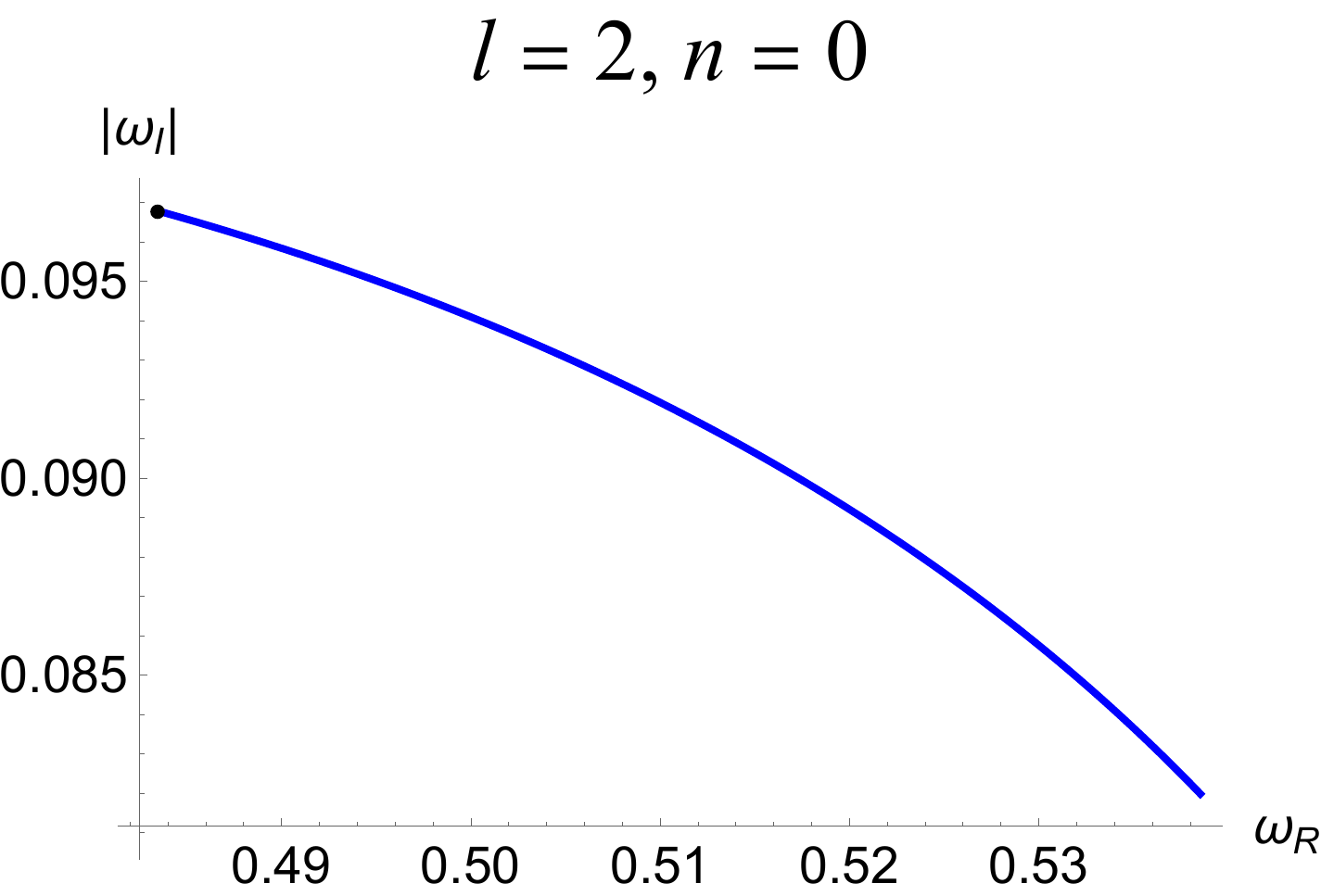}\hspace{1mm}
		\includegraphics[width=5.2cm]{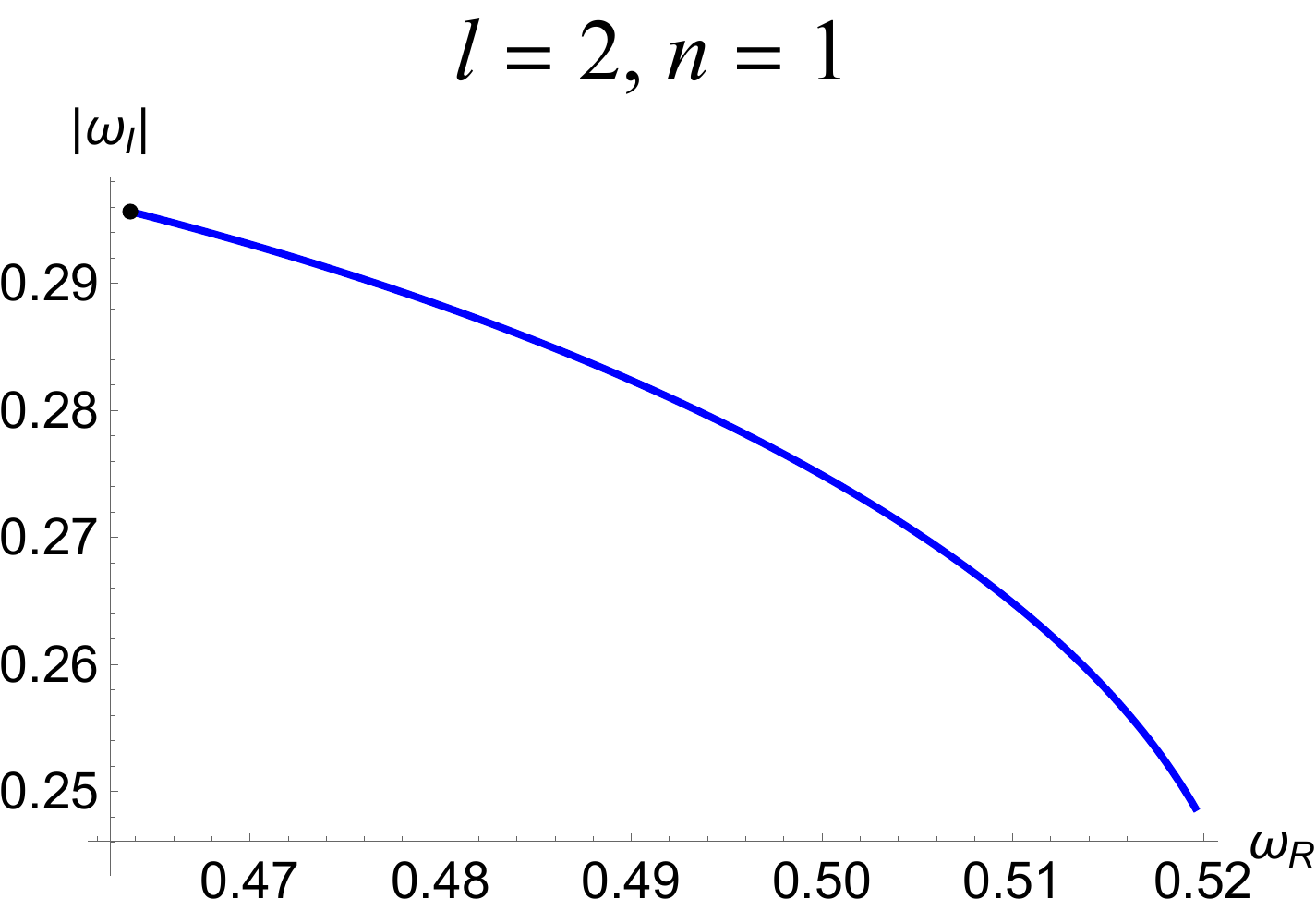}\hspace{1mm}
		\includegraphics[width=5.2cm]{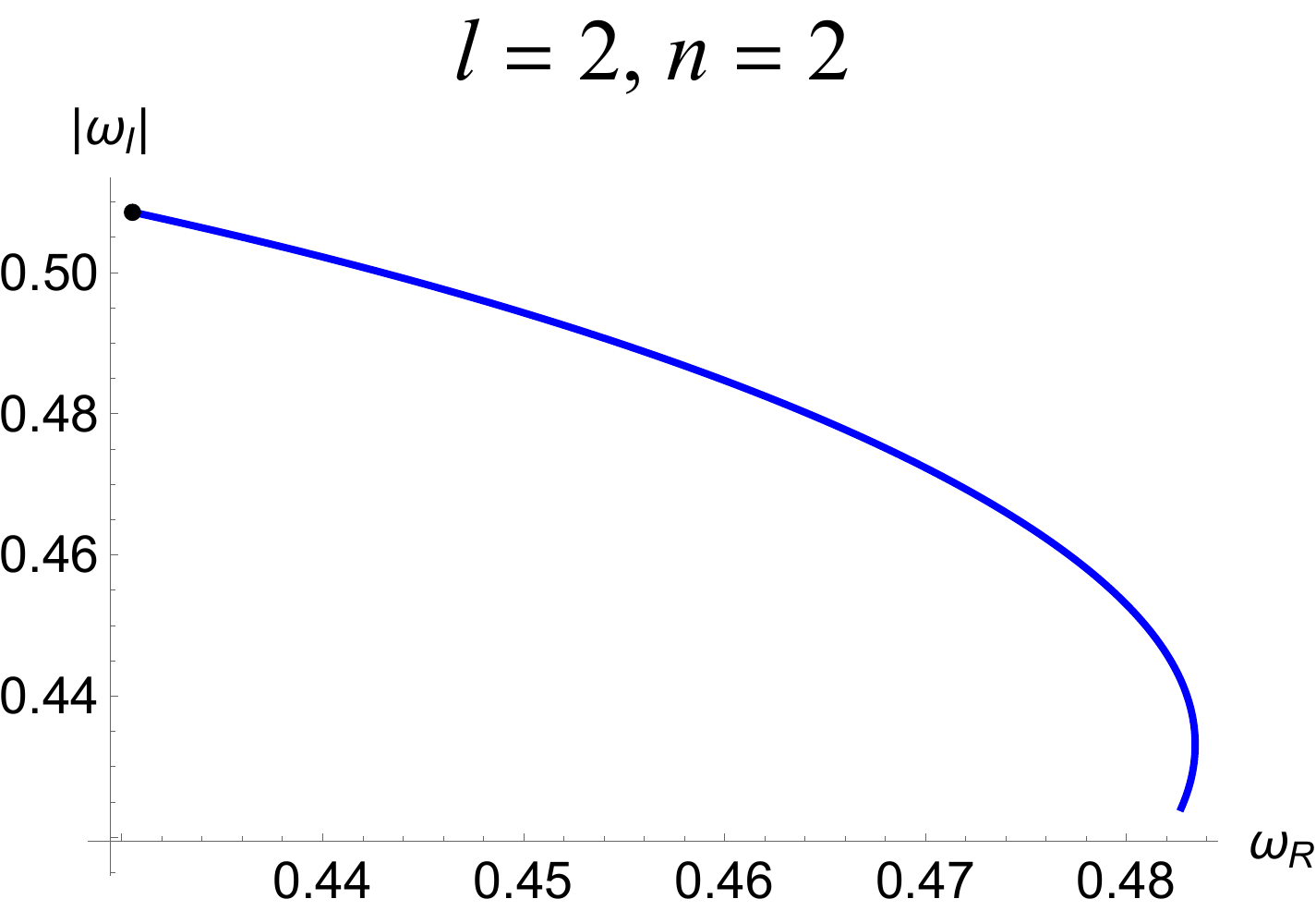}\hspace{1mm}
		\includegraphics[width=5.2cm]{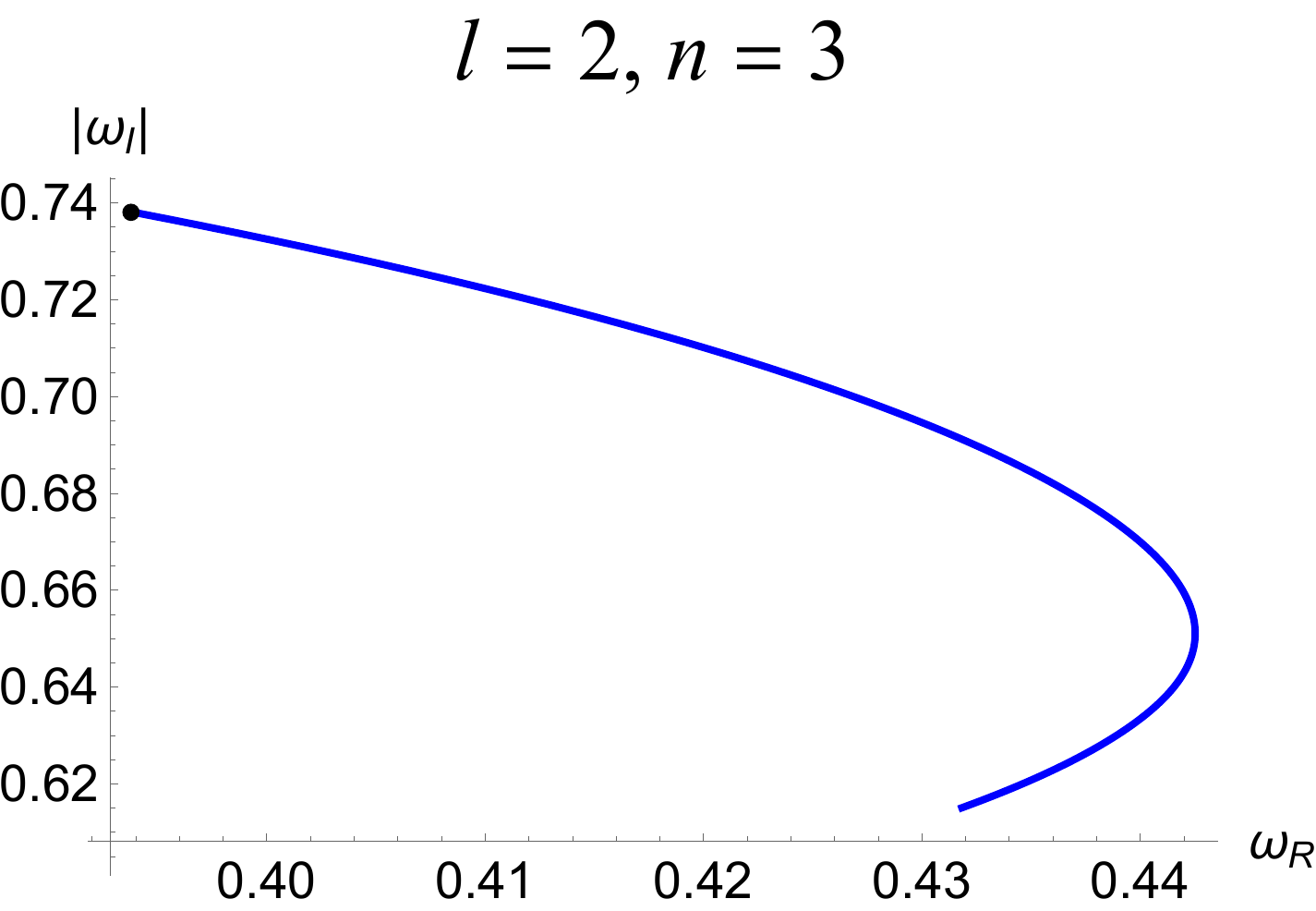}\hspace{1mm}
		\includegraphics[width=5.2cm]{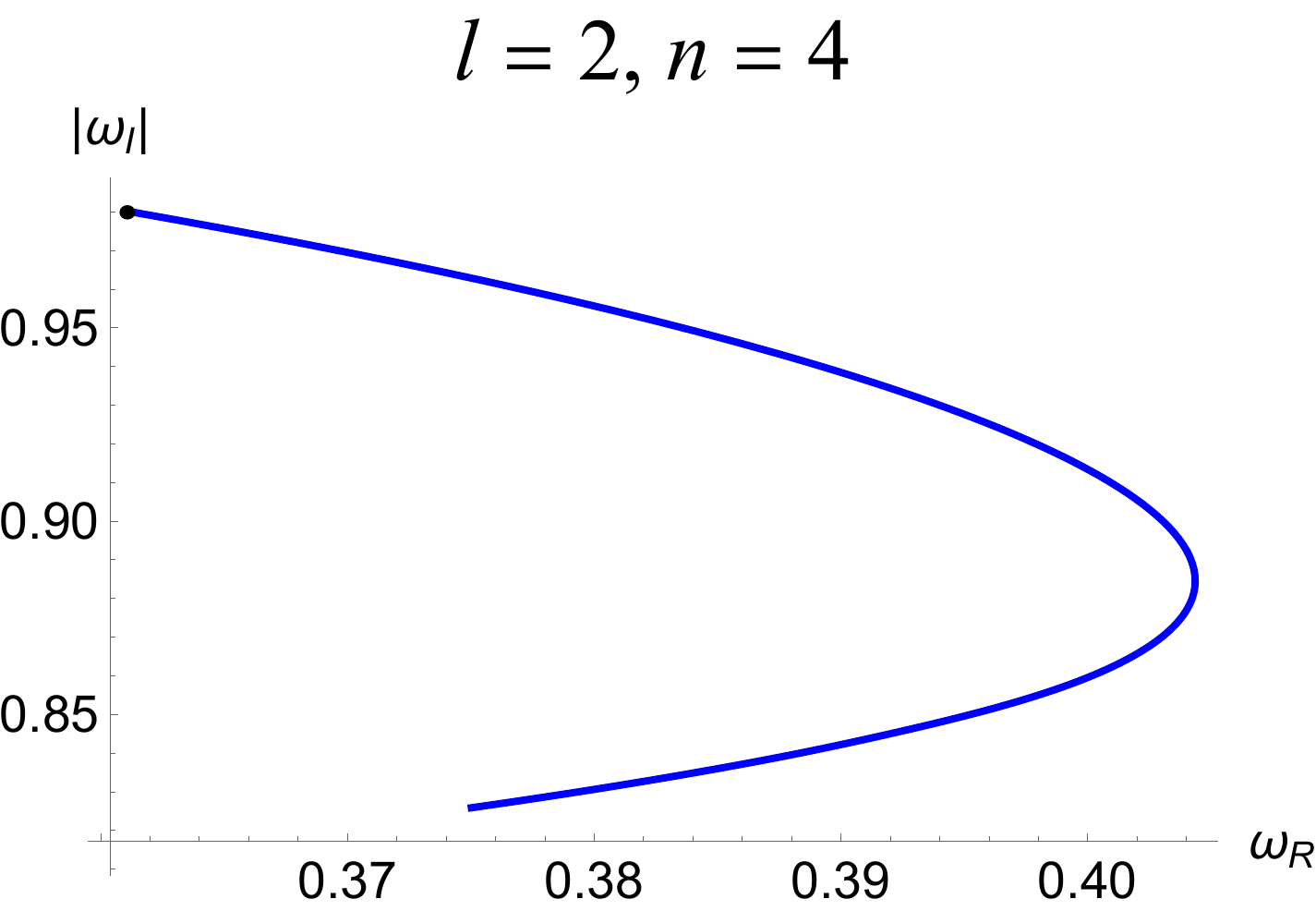}\hspace{1mm}
		\includegraphics[width=5.2cm]{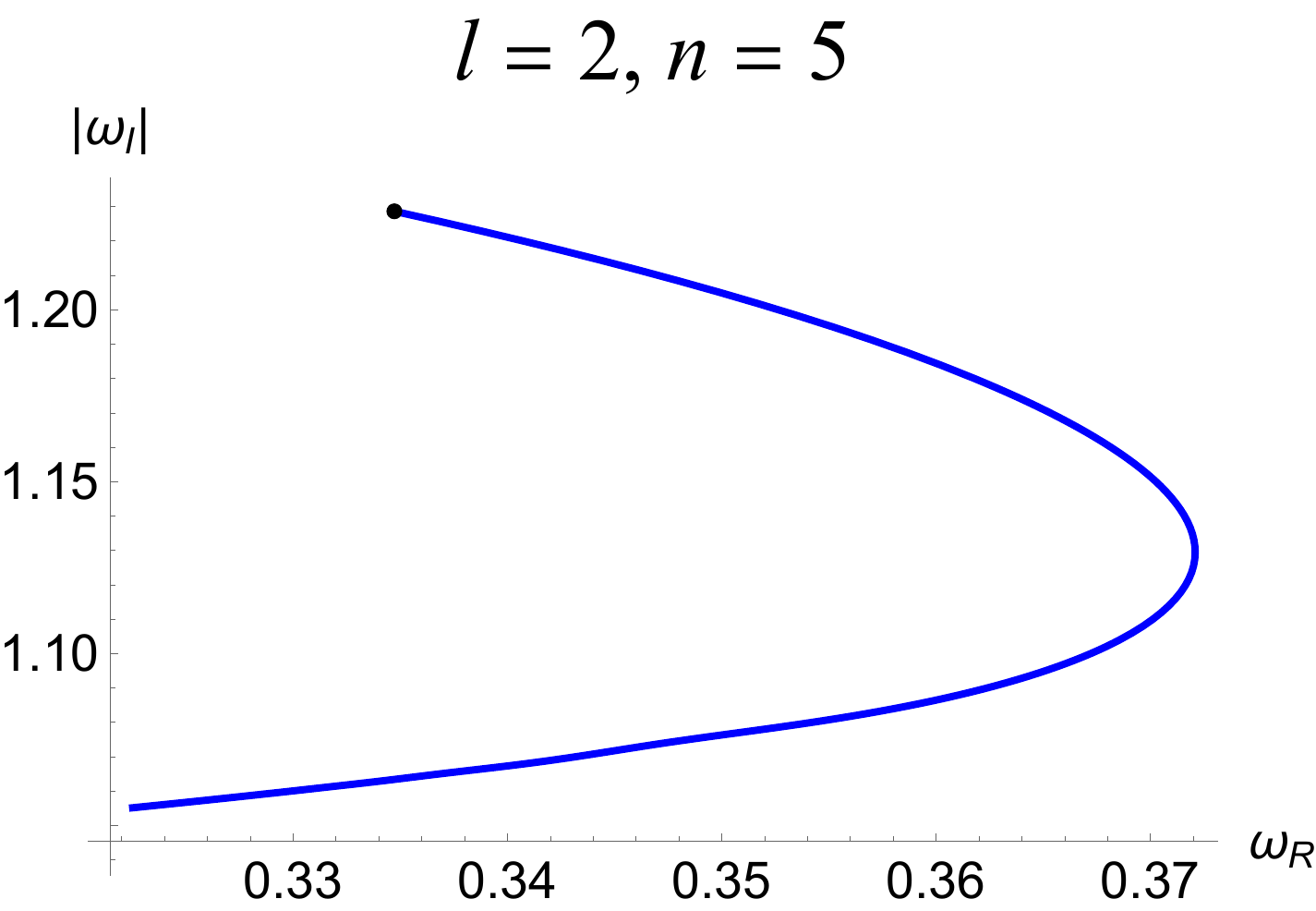}\hspace{1mm}
		\caption{The trajectory of the QNMs with the variation of $\alpha_0$ on the frequency plane with $l=2$} under scalar field perturbation.
		\label{l=2,n=0，1，2，3,4,5}
	}
\end{figure}

The feature of the QNMs demonstrated in these figures  can be summarized as follows,

1. Firstly, once $\alpha_0$ and $l$ are fixed, the real part of the QNM frequency decreases with the increase of  the overtone number $n$, while the imaginary part increases, which directly leads to a faster decay of the QNMs.

2. When $\alpha_0$ and $n$ are fixed, the increase of the angular quantum number $l$ leads to the increase of the real part of the frequency, while the imaginary part remains almost unchanged. This is because the effective potential increases with $l$, and it requires that the wave acquires greater energy to propagate to infinity.

3. With the running of $\alpha_0$, the QNMs frequency intend to exhibit a non-monotonic behavior and spiral behavior. When $l$ is fixed at $l=0, 1$, as the overtone number $n$ increases, the non-monotonic behavior becomes more evident and may evolve into a trajectory with a spiral behavior, for instance in figures for $(l=0,n=1)$, $(l=1,n=4)$ and $(l=1,n=5)$. However, an increase in the angular momentum quantum number $l$ seems to suppress this non-monotonic behavior. For instance, when $l=2$, the non-monotonic behavior is not evident until $n>2$ and the spiral behavior of the trajectory does not appear in all the figures with $n\leq 5$. 
\begin{figure}[ht]
		\centering{
			\includegraphics[width=7.8cm]{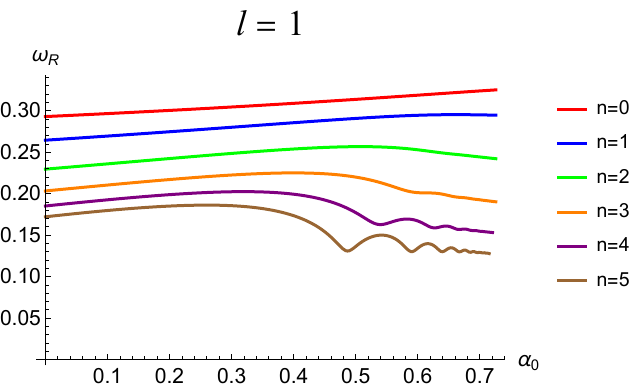}\hspace{6mm}
			\includegraphics[width=7.8cm]{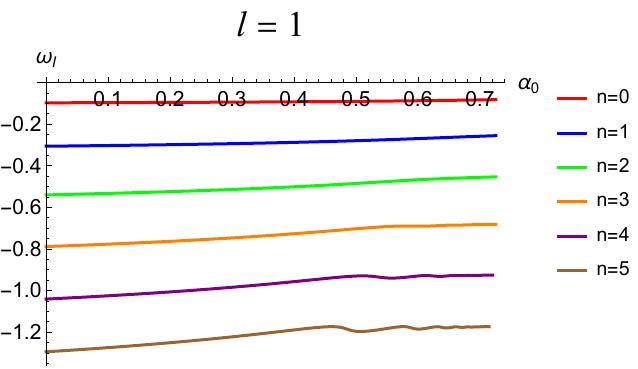}
			\caption{The real and imaginary parts of QNMs as the function of $\alpha_0$ with $l=1$ under scalar field perturbation.}
			\label{Real part Imaginary part l=1}
		}
\end{figure}
\begin{figure}[ht]
		\centering{
			\includegraphics[width=7.8cm]{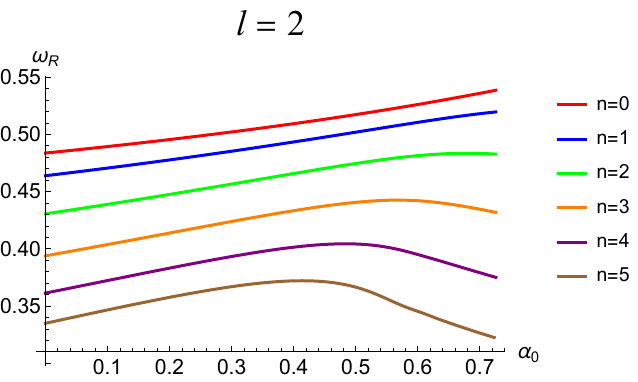}\hspace{6mm}
			\includegraphics[width=7.8cm]{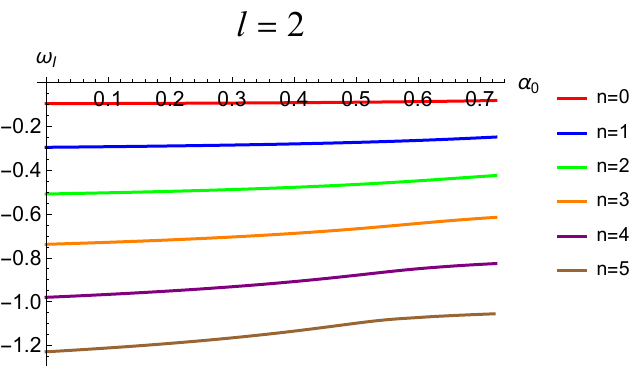}
			\caption{The real and imaginary parts of QNMs as the function of $\alpha_0$ with $l=2$ under scalar field perturbation.}
			\label{Real part Imaginary part l=2}
		}
\end{figure}

To manifestly demonstrate the change of the QNMs frequency with the parameter $\alpha_0$, we may directly plot the real part and the imaginary part of the frequency as the function of $\alpha_0$ separately, as illustrated in FIG. \ref{Real part Imaginary part l=1} for $l=1$ and FIG. \ref{Real part Imaginary part l=2} for $l=2$. It is evident that when $l=1$ and the value of $n$ reaches to 4 and 5, the real part exhibits significant oscillatory behavior, known as the outburst phenomenon. The imaginary part also exhibits the outburst phenomenon in the cases of $n=4$ and $n=5$. Furthermore, the higher the overtone number $n$, the more pronounced the oscillatory behavior becomes in both the real and imaginary parts. These oscillations in the real and imaginary parts contribute to the spiral behavior observed in the previous figures. However, when $l=2$, the oscillatory behavior becomes weak and disappearing, as shown in FIG. \ref{Real part Imaginary part l=2}. In addition, the real part only exhibits the non-monotonic behavior for $n=3$, $n=4$ and $n=5$, while the imaginary part remains almost unchanged. It is expected to confirm whether a higher overtone number $n$ would cause the non-monotonic behavior and further evolve into oscillatory behavior in future. 

\subsection{The QNMs under the perturbation of electromagnetic field}
\begin{figure}[ht]
		\centering{
			\includegraphics[width=7.8cm]{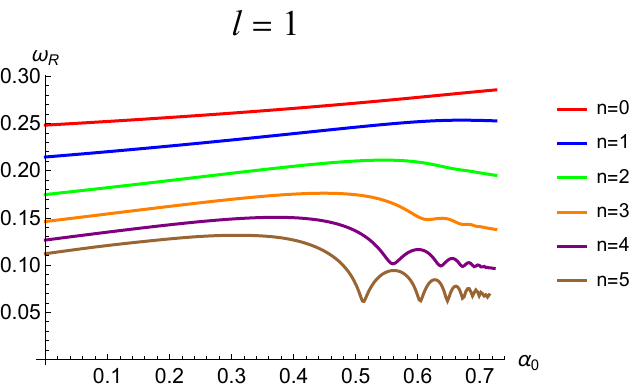}\hspace{6mm}
			\includegraphics[width=7.8cm]{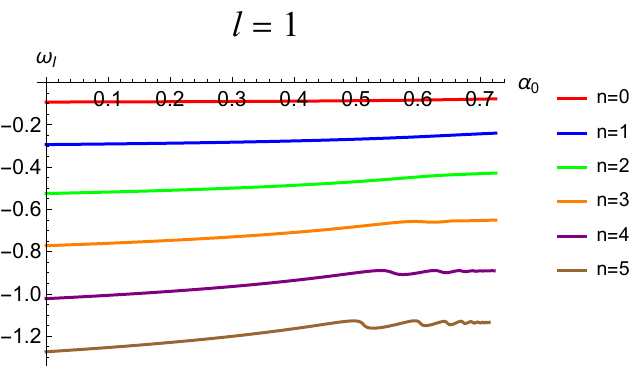}
			\caption{The real and imaginary parts of QNMs as the function of $\alpha_0$ with $l=1$ under electromagnetic field perturbation.}
			\label{spin1 Real part Imaginary part l=1}
		}
\end{figure}

\begin{figure}[ht]
		\centering{
			\includegraphics[width=7.8cm]{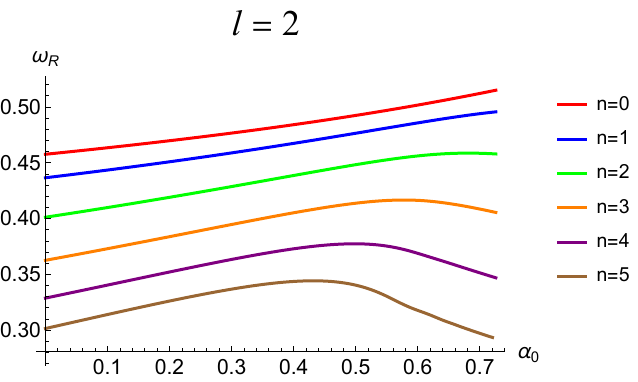}\hspace{6mm}
			\includegraphics[width=7.8cm]{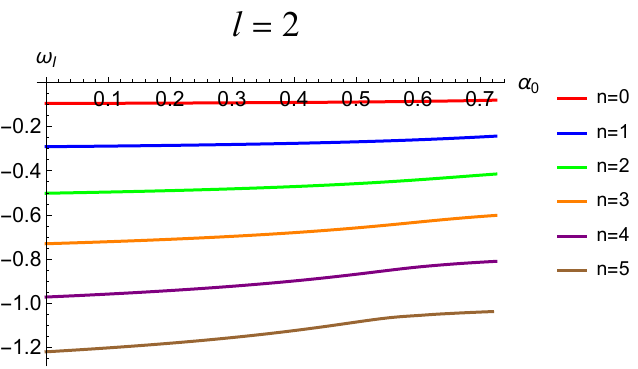}
			\caption{The real and imaginary parts of QNMs as the function of $\alpha_0$ with $l=2$ under electromagnetic field perturbation.}
			\label{spin1 Real part Imaginary part l=2}
		}
\end{figure}

In this subsection we continue to investigate the response of the regular black hole under electromagnetic field perturbations.
In parallel, the QNMs of the regular black hole under electromagnetic field perturbations can be computed for different values of $l, n, \alpha_0$, which have been summarized in Tables (\ref{ele l=1}), (\ref{ele l=2}) and FIG. \ref{spin1 l=1,n=0,1,2,3,4,5}, \ref{spin1 l=2,n=0,1,2,3,4,5}, \ref{spin1 Real part Imaginary part l=1} and \ref{spin1 Real part Imaginary part l=2}. 
 
\begin{table}[ht]
\centering
\begin{tabular}{c@{\hspace{1em}}c@{\hspace{2em}}c@{\hspace{2em}}c@{\hspace{2em}}c} 
\hline
\hline
$n$ & $\alpha_0=0$ & $\alpha_0=0.3$ & $\alpha_0=0.5$ & $\alpha_0=2/e$  \\ 
\hline
0 &0.248263-0.092488i& 0.261020-0.089823i &0.271566-0.086200i &0.285600-0.077192i  \\
1 &0.214515-0.293668i&0.232596-0.282407i &0.246087-0.268233i & 0.252756-0.239495i \\
2 &0.174774-0.525188i& 0.197380-0.499535i &0.210289-0.469126i &0.195035-0.428313i\\
3 &0.146176-0.771909i& 0.169757-0.730175i &0.174524-0.681999i &0.137729-0.650895i\\
4 &0.126548-1.022530i& 0.148845-0.964811i &0.136150-0.900198i&0.096489-0.890878i\\
5 &0.112103-1.273890i &0.131601-1.200420i& 0.079902-1.127980i& 0.068540-1.134290i\\
\hline
\hline
\end{tabular}
\caption{The frequency of QNMs under electromagnetic perturbations for $l = 1$.}
\label{ele l=1}
\end{table}
\begin{table}[ht]
\centering
\begin{tabular}{c@{\hspace{1em}}c@{\hspace{2em}}c@{\hspace{2em}}c@{\hspace{2em}}c} 
\hline
\hline
$n$ & $\alpha_0=0$ & $\alpha_0=0.3$ & $\alpha_0=0.5$ & $\alpha_0=2/e$  \\ 
\hline
0 &0.457596-0.095004i&0.476647-0.092046i&0.492386-0.088424i&0.514945-0.080055i \\
1 &0.436542-0.290710i&0.458901-0.280724i&0.476425-0.268720i &0.495597-0.242907i \\
2 &0.401187-0.501587i&0.428766-0.481513i&0.448256-0.458131i&0.457868-0.414559i\\
3 &0.362595-0.730199i&0.394796-0.696651i&0.413739-0.658717i&0.405478-0.601951i\\
4 &0.328737-0.971609i&0.363491-0.922730i&0.377384-0.868702i&0.346929-0.810019i\\
5 &0.301493-1.21971i & 0.336867-1.154990i& 0.340336-1.085250i&0.293947-1.037730i \\
\hline
\hline
\end{tabular}
\caption{The frequency of QNMs under electromagnetic perturbations for $l = 2$.}
\label{ele l=2}
\end{table}

\begin{figure}[ht]
    \centering
    \includegraphics[width=5.2cm]{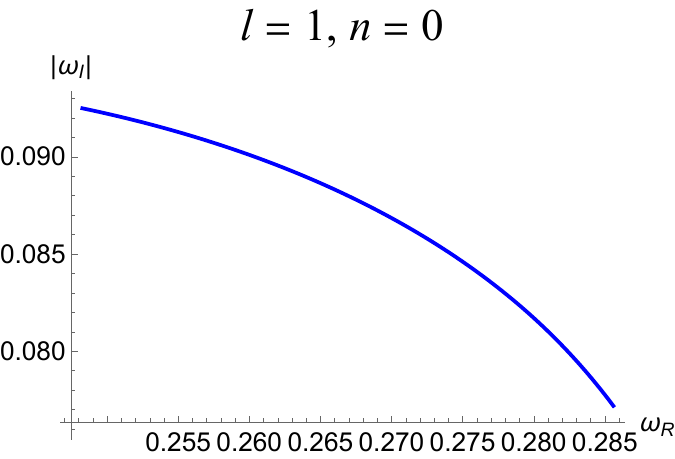}\hspace{1mm}
    \includegraphics[width=5.2cm]{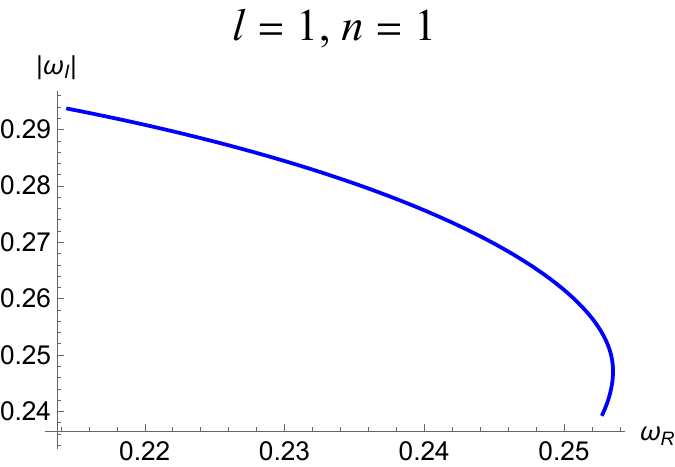}\hspace{1mm} 
    \includegraphics[width=5.2cm]{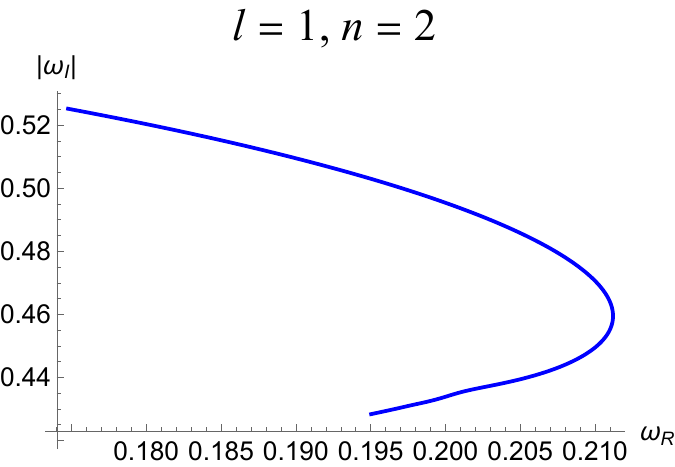}\hspace{1mm} 
    \includegraphics[width=5.2cm]{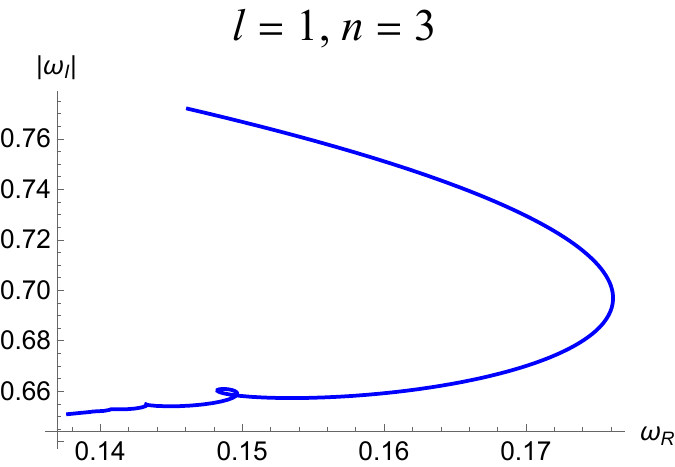}\hspace{1mm}
    \includegraphics[width=5.2cm]{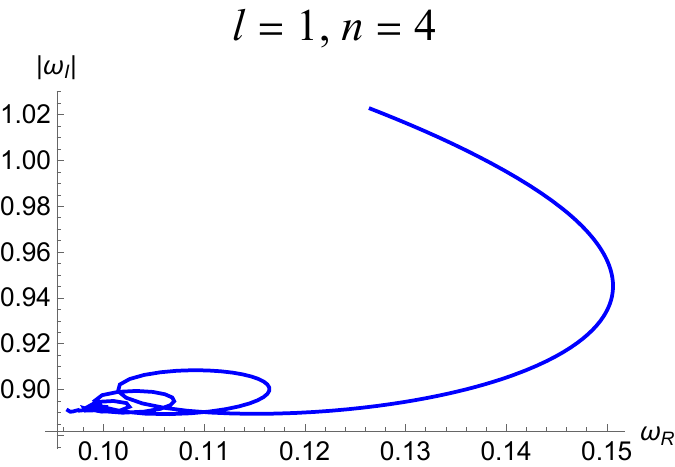}\hspace{1mm} 
    \includegraphics[width=5.2cm]{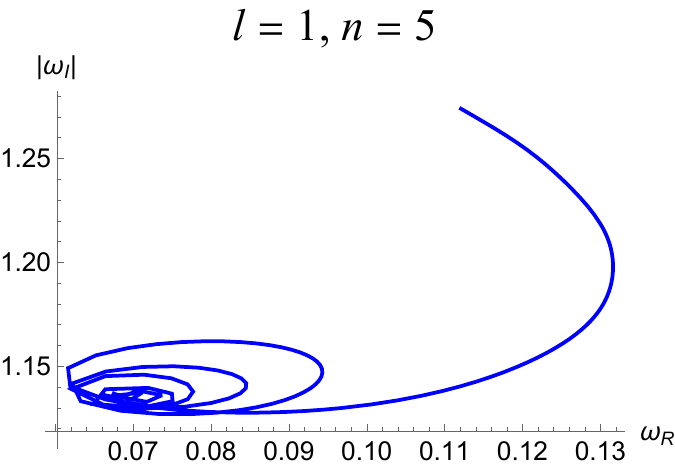}\hspace{1mm} 
    \caption{The trajectory of the QNMs with the variation of $\alpha_0$ on the frequency plane with $l=1$ under electromagnetic field perturbation}
    \label{spin1 l=1,n=0,1,2,3,4,5}
\end{figure}

\begin{figure}[ht]
    \centering
    \includegraphics[width=5.2cm]{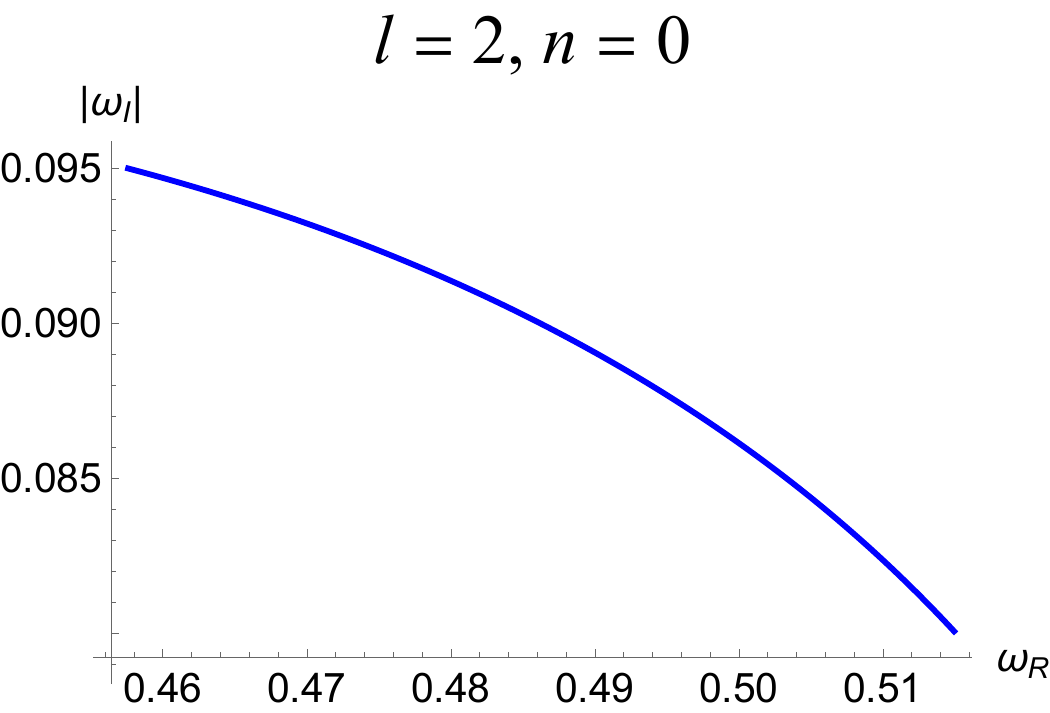}\hspace{1mm}
    \includegraphics[width=5.2cm]{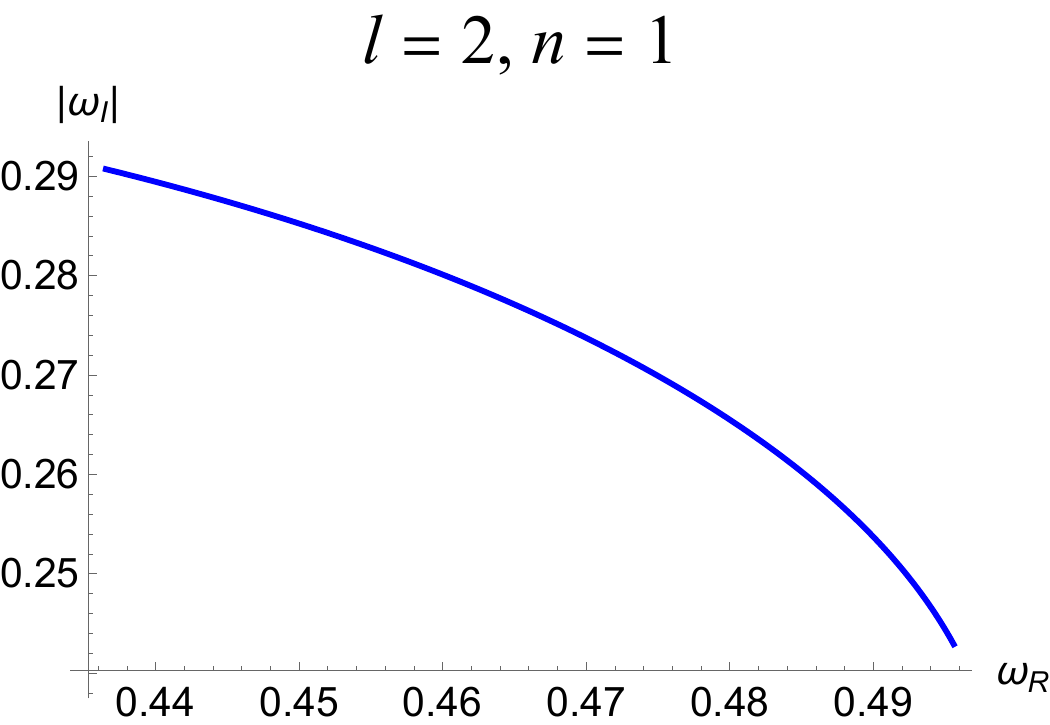}\hspace{1mm} 
    \includegraphics[width=5.2cm]{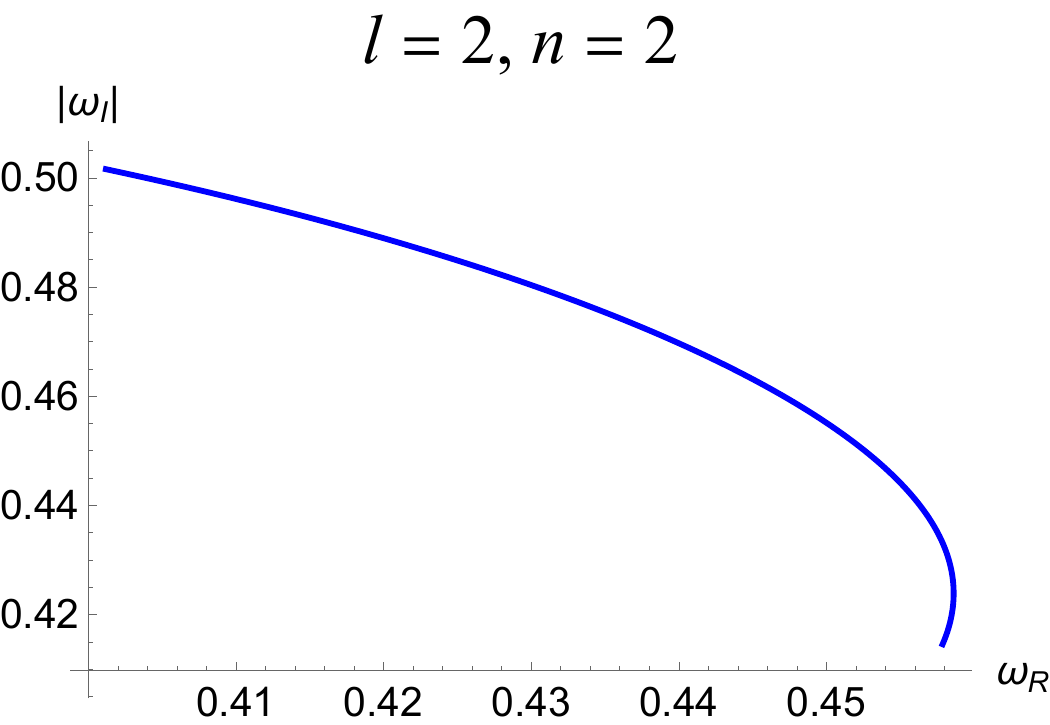}\hspace{1mm} 
    \includegraphics[width=5.2cm]{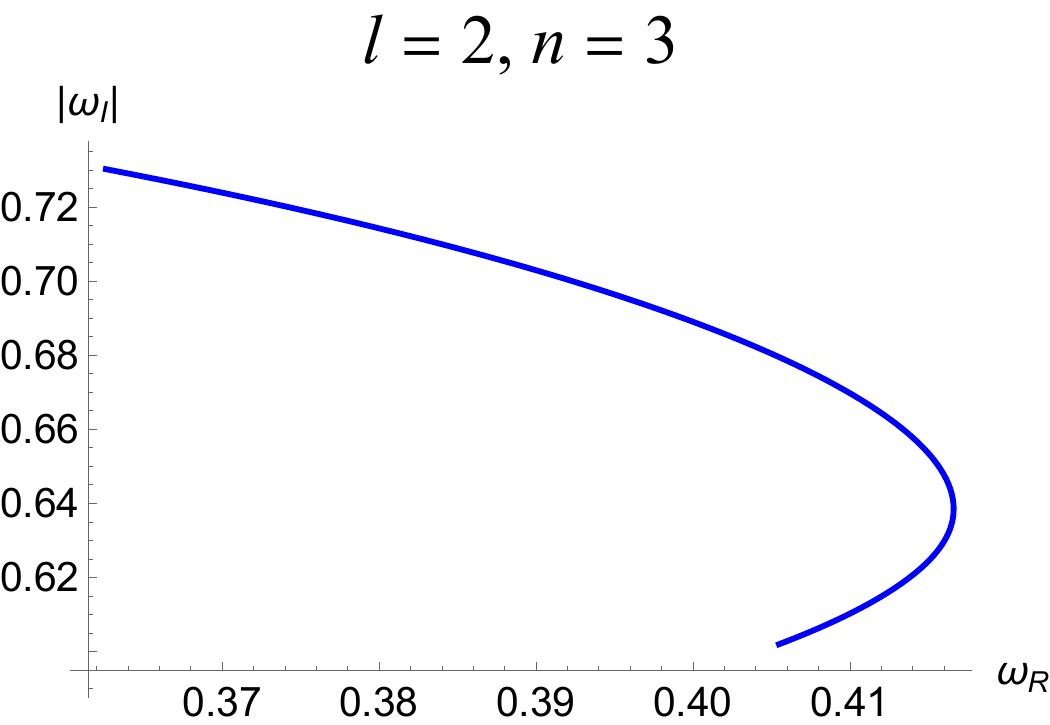}\hspace{1mm}
    \includegraphics[width=5.2cm]{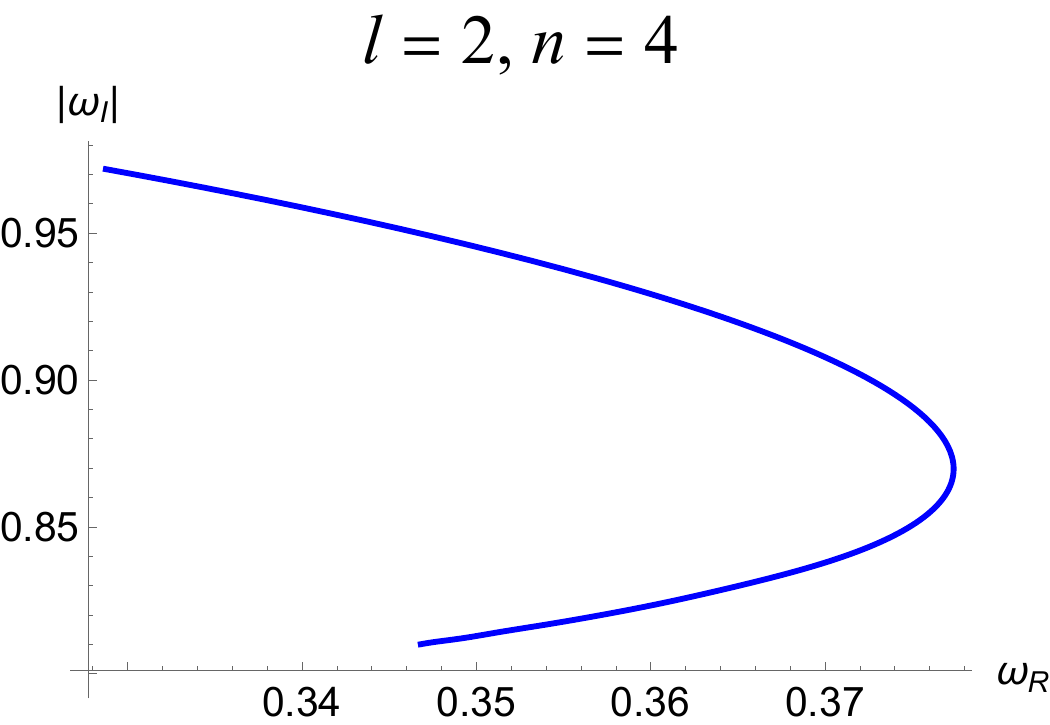}\hspace{1mm} 
    \includegraphics[width=5.2cm]{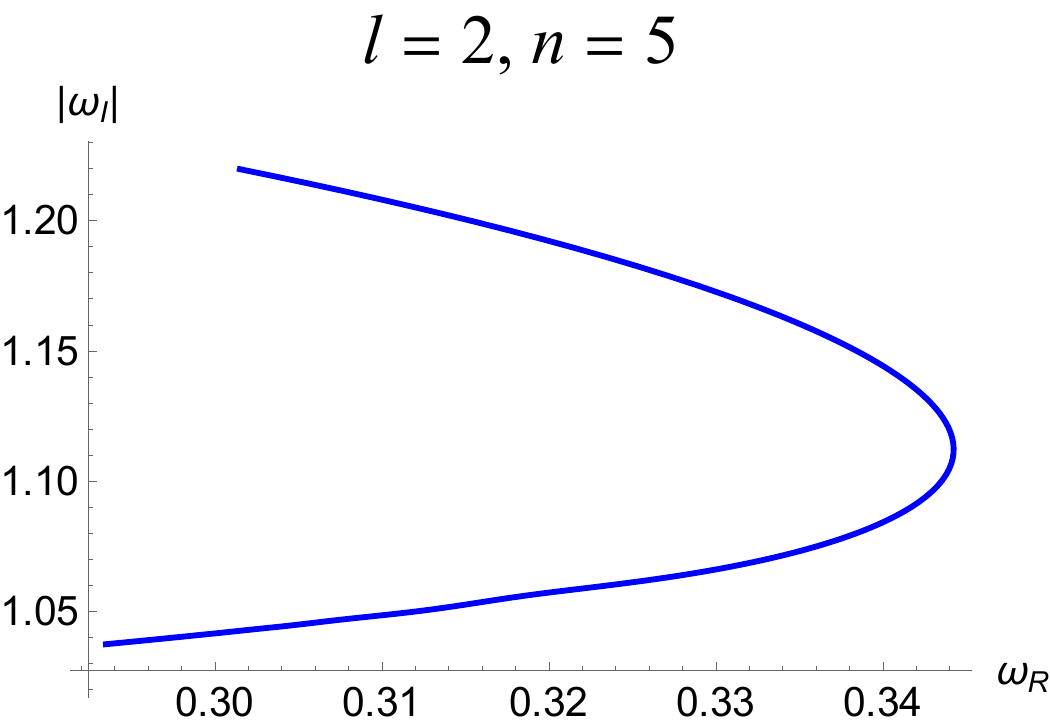}\hspace{1mm} 
    \caption{The trajectory of the QNMs with the variation of $\alpha_0$ on the frequency plane with $l=2$ under electromagnetic field perturbation.}
    \label{spin1 l=2,n=0,1,2,3,4,5}
\end{figure}

We notice that the QNMs under electromagnetic field perturbation exhibit the similar behavior as those under the scalar field perturbation. 
In particular, it can be seen from FIG. \ref{spin1 l=1,n=0,1,2,3,4,5} and FIG. \ref{spin1 l=2,n=0,1,2,3,4,5} that for $l=1$, the spiral behavior can be obviously observed with the increase of $n$, while for $l=2$, the behavior disappears. Qualitatively this observed phenomenon is similar to that in the case of the scalar field perturbation. Furthermore,  the occurrence of  the spiral behavior on the plane of the complex frequency can be reflected by the oscillatory behavior of the real and imaginary parts of the QNMs as the function of the deviation parameter $\alpha_0$, as illustrated in FIG. \ref{spin1 Real part Imaginary part l=1} and \ref{spin1 Real part Imaginary part l=2}. 

In addition, in comparison with the QNMs under the scalar field perturbation, we find that the QNMs with the same parameters under electromagnetic field perturbation have smaller real and imaginary parts. This suggests that the QNMs under electromagnetic field perturbation have a lower oscillation frequency, as well as a weaker attenuation.

\section{The comparison of the QNMs between this regular  black hole  and Bardeen black hole  }
\label{Comparison between QNMs of regular BHs and Bardeen BHs}

As we mentioned in the previous section, the regular black hole with $(x=2/3, c=2)$ has the same asymptotic behavior at infinity as the Bardeen black hole. The difference lies on the center of the black hole, as the former has a Minkowskian core while the latter has a de-Sitter core. It is quite interesting to compare the QNM behavior of these two regular black holes, that is what we intend to investigate in this section. Previously the QNMs of Bardeen black hole have been investigated in \cite{Fernando:2012yw,Toshmatov:2015wga,Saleh:2018hba,Jusufi:2020agr,Sun:2023woa,Konoplya:2023ahd}. Here, we present the numerical results with the PS method, with a focus to compare its behavior with that of the regular black hole with $(x=2/3, c=2)$. 

We investigate the QNMs of Bardeen black hole and define the discrepancy of the QNM frequency between Bardeen black hole and the regular black hole with $(x=2/3, c=2)$ as
\begin{equation}
	\Delta=\left |\frac{\omega_{Regular}-\omega_{Bardeen}}{\omega_{Regular}}\right | \times 100\%\,.
	\label{errorv2}
\end{equation}

Table (\ref{Comparison scalar}) lists the QNMs of Bardeen BHs and  the regular BH with with $(x=2/3, c=2)$ under the scalar field perturbation, while Table (\ref{Comparison electromagnetic}) lists the QNMs  under the electromagnetic field perturbation.
\begin{table}[ht]
\centering
\begin{tabular}{@{\hspace{0.5em}}c@{\hspace{0.5em}}@{\hspace{0.5em}}c@{\hspace{0.5em}}@{\hspace{1em}}c@{\hspace{1em}}@{\hspace{1em}}c@{\hspace{1em}}c} 
\hline\hline
$\alpha_0$& \centering $n$& \centering$\omega_{Regular}$\centering & $\omega_{Bardeen}$ & $\Delta$ \\ 
\hline
\multirow{6}{*}{0.15} & 0 &  0.298396-0.096297i & 0.298395-0.096298i &  4.5 $\times$ $10^{-4}\%$ \\
                  & 1 & 0.271993-0.300992i &  0.271994-0.300997i &  0.0013$\%$  \\
                  & 2 & 0.239121-0.528880i &  0.239127-0.528891i &  0.0022$\%$  \\
                  & 3 & 0.213779-0.770343i &  0.213795-0.770359i &  0.0028$\%$ \\
                  & 4 & 0.195902-1.016091i &  0.195933-1.016112i &  0.0036$\%$\\
                  & 5& 0.182764-1.262781i  &  0.182814-1.262811i &  0.0046$\%$\\ 
\hline
\multirow{6}{*}{0.58} & 0& 0.317685-0.088317i&  0.317789-0.088508i&  0.066$\%$\\
	 & 1& 0.293929-0.271799i&  0.294912-0.272364i&  0.28$\%$ \\
	 & 2& 0.254437-0.470637i&  0.258373-0.470546i&  0.74$\%$\\
	 & 3& 0.203826-0.689841i&  0.213611-0.681019i&  1.84$\%$\\
	 & 4& 0.169328-0.935988i&  0.152653-0.910328i&  3.21$\%$\\
	 & 5& 0.135243-1.173540i&  0.114004-1.176200i&  1.81$\%$\\
\hline\hline
\end{tabular}
\caption{A comparison of the QNMs under scalar field perturbation between the regular BH with $(x=2/3, c=2)$ and the Bardeen BH. Here, we have fixed $l=1$.}
\label{Comparison scalar}
\end{table}

\begin{table}[ht]
\centering
\begin{tabular}{@{\hspace{0.5em}}c@{\hspace{0.5em}}@{\hspace{0.5em}}c@{\hspace{0.5em}}@{\hspace{1em}}c@{\hspace{1em}}@{\hspace{1em}}c@{\hspace{1em}}c} 
\hline
\hline
$\alpha_0$&  $n$& $\omega_{Regular}$ & $\omega_{Bardeen}$ & $\Delta$ \\ 
\hline
\multirow{6}{*}{0.15} & 0& 0.254697-0.091332i&  0.25479 - 0.091242 i&   7.4$\times$$10^{-4}$ $\%$\\
	 & 1& 0.223744-0.288622i&  0.223744-0.288629i&   0.0019$\%$ \\
	 & 2& 0.186622-0.513412i&  0.186626-0.513428i&   0.0030$\%$\\
	 & 3& 0.159152-0.752567i&  0.159167-0.752592i&   0.0038$\%$\\
	 & 4& 0.139809-0.995621i&  0.139324-0.993985i&   0.0047$\%$\\
	 & 5& 0.125218-1.23947i&  0.12527-1.239520i&   0.0056$\%$\\ 
\hline
\multirow{6}{*}{0.58} & 0& 0.276441-0.083850i&  0.276483-0.084136i&  0.10$\%$\\
	 & 1& 0.250942-0.259415i&  0.251826-0.260473i&  0.38$\%$ \\
	 & 2& 0.210432-0.451302i&  0.214505-0.452913i&  0.88$\%$\\
	 & 3& 0.158754-0.658531i&  0.172191-0.656379i&  2.00$\%$\\
	 & 4& 0.111666-0.907950i&  0.118232-0.866031i&  4.64$\%$\\
	 & 5& 0.089483-1.133720i&  0.046411-1.131700i&  3.79$\%$\\
\hline\hline
\end{tabular}
\caption{A comparison of the QNMs under electromagnetic field perturbation between the regular BH with $(x=2/3, c=2)$ and the Bardeen BH. Here, we have fixed $l=1$.}
\label{Comparison electromagnetic}
\end{table}
 Some similar phenomena have been observed in both cases. Without loss of generality, we fix $l=1$ and compare the QNMs with various $n$. Firstly, we notice that with the same deviation parameter $\alpha_0$, the discrepancy of the fundamental mode with $n=0$ is the minimal, while with the increase of $n$, the difference of the frequency becomes larger. Secondly, as the  deviation parameter $\alpha_0$ becomes larger, the overall discrepancy of the frequency becomes larger as well, implying that one could distinguish these two black holes easier. This tendency is understandable since the difference of the corresponding effective potential of these two black holes becomes more evident with the increase of $\alpha_0$.   

\section{The comparison of the QNMs of this sort of regular BHs with different parameters X and n}
\label{Comparison of QNMs of regular BHs with different parameters X and n under scalar field perturbations}

As we mentioned before, a sort of regular black holes with sub-Planck Kretschmann curvature can be constructed with the different choice of the parameter $x$ and $c$, under the condition $c \geq x \geq c/3$ and $c\geq 2$ \cite{ling2023regular}. Specially,  $(x=2/3,c=2)$ and  $(x=1,c=3)$ are just two typical regular black holes which have the identical asymptotic behavior at infinity as the famous Bardeen black hole and Hayward black hole, respectively. In this section, we intend to investigate the QNMs of this sort regular black holes and compare their behavior with different values of $x$ and $n$.  

We plot the trajectory of the QNMs on the plane of the frequency as the deviation parameter $\alpha_0$ varies, where FIG. \ref{Different l=0,n=0} is for $l=0$, $n=0$ and FIG. \ref{Different l=0,n=1} is for $l=0$, $n=1$.

Firstly, we look at FIG. \ref{Different l=0,n=0} for the fundamental mode of the regular black holes with different $x$ and $c$. In general, each trajectory starts from the same point which is identified as the frequency of Schwarzschild black hole, and exhibits the non-monotonic behavior with the similar shape. As $\alpha_0$ runs from zero to the maximal value, the magnitude of the imaginary part is always smaller than the imaginary part of Schwarzschild black hole, no matter what values $x$ and $c$ are. Secondly, we notice that the trajectory moves toward the left as the parameter $c$ increases, indicating that the non-monotonic behavior becomes pronounced. As a result, the maximal value of the real part of the frequency becomes smaller as $c$ increases and the real part with the maximal $\alpha_0$, which is the ending point of the trajectory, may be smaller than that of Schwarzschild black hole, as illustrated for $c=4$. Thirdly, we notice that the trajectory of  the regular black hole with ($x=2/3, c=2$) is overlapped with the trajectory of  the regular black hole with ($x=1, c=2$), implying it is independent of the parameter $x$. This results from the fact that  in this setup we have set $M=1$, thus the  effect of $x$ is absent. Nevertheless, we intend to argue that all the figures should be independent of the setup of Mass $M$, once one appropriately chooses the dimensionless quantity as the coordinate. For instance, taking $\omega_RM$ and $|\omega_I|M$  as the unit of the axes, then the figure should be independent of the setup of $M$.  
\begin{figure}[ht]
	\centering{
		\includegraphics[width=10cm]{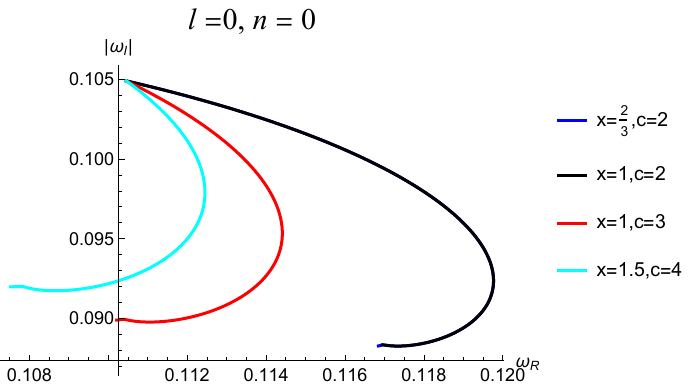}
		\caption{The trajectory of the QNMs with the variation of $\alpha_0$ on the frequency plane with  $l=0$ and $n=0$ for regular black holes with various values of $x$ and $c$.}
  \label{Different l=0,n=0}
  }
\end{figure}

\begin{figure}[ht]
	\centering{
		\includegraphics[width=10cm]{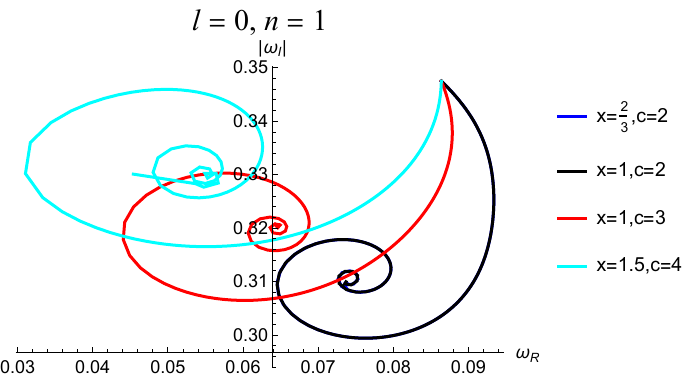}
		\caption{The trajectory of the QNMs with the variation of $\alpha_0$ on the frequency plane with  $l=0$ and $n=1$ for regular black holes with various values of $x$ and $c$.}
            \label{Different l=0,n=1}
            }
\end{figure}
Next we turn to FIG. \ref{Different l=0,n=1} for ($l=0$ and $n=1$). In this case one notices that all the trajectories exhibit a spiral behavior, implying that the appearance of the spiral behavior of QNMs is only determined by the principal quantum number $n$ and the angular quantum number $l$, irrespective of the parameter $x$ and $c$. Secondly, for larger  $c$, the real part of the QNMs decreases while the imaginary part increases such that the trajectory of the frequency moves to the left, indicating that the oscillation frequency of the QNM becomes weaker but the attenuation increases.  

\section{CONCLUSION AND DISCUSSION}
\label{CONCLUSION AND DISCUSSION}
In this paper we have computed the QNMs of a sort of regular black holes which is characterized by a Minkowskian core and sub-Planckian curvature with pseudo-spectral method and investigated the feature of these QNMs under the perturbations of scalar field and electromagnetic field, respectively. As a typical example,  we have focused  on the specific regular BH with fixed parameters $x = 2/3$ and $c = 2$, which has the same asymptotic behavior at infinity with Bardeen black hole. 
It turns out that on the plane of the complex frequency, with the increase of $n$, the trajectory of the QNMs intends to exhibit a non-monotonic behavior, and then turn to a spiral behavior. For larger $l$, the value of $n$ leading to spiral behavior becomes larger as well. At high overtones, the oscillation of the imaginary and real parts is referred to as the outburst of the overtone. We have also compared the QNMs of this regular black hole with those of the Bardeen BH.  It is found that when the deviation parameter $\alpha_0$ and the principal quantum number $n$ are increased, the discrepancy  between the QNMs of these two black holes  becomes larger as well. Therefore, QNMs is undoubtedly a useful tool for distinguishing these BHs.  Finally, we have compared the QNMs of the regular black holes with different $x$ and $c$. We have found that the values of $c$ and $x$ do not change the non-monotonic and spiral behavior qualitatively, but with the increase of  $c$, the real part of QNM decreases and the imaginary part increases overall, implying slower oscillations and stronger attenuation. Therefore, QNMs are also an effective tool to distinguish  the  regular BHs with different parameters $x$ and $c$.\par

\section*{Acknowledgments}
We are very grateful to Zhong-wen Feng, Pan Li, Kai Li, Peng Liu, Wen-bin Pan, Meng-he Wu and Zhangping Yu  for helpful discussions. Specially we would like to thank Wen-bin Pan,Meng-he Wu and Zhangping Yu for improving the code on the numerical simulation with PS method. This work is supported in part by the Natural Science Foundation
of China under Grant No.~12035016 and 12275275. It is also supported by Beijing Natural Science Foundation under Grant No. 1222031, the Innovative Projects of Science and Technology No. E2545BU210 at IHEP, by Sichuan Youth Science and Technology Innovation Research Team with Grant No.~21CXTD0038, by and the National Natural Science Foundation of Sichuan Province with Grant No. 2023NSFSC1352, by The Central Guidance on Local Science and Technology Development Fund of Sichuan Province (2024ZYD0075),  by the Sichuan Science and Technology Program (2023ZYD0023), and by Sichuan Science and Technology Program (2024NSFSC1999).

\appendix
\section{Pseudo-spectral method}
\label{sec:pseudospectral-method}

The pseudo-spectral method as a highly efficient numerical method is widely applied to solve various eigenvalue problems\cite{dias2009instability,dias2010instability,destounis2021pseudospectrum,jaramillo2022gravitational,monteiro2010semiclassical,xiong2022quasinormal}. It also has an advantage in determining the QNMs of high overtones\cite{zhang2024quasinormal,fu2023quasinormal,fu2024peculiar,jansen2017overdamped}. The key in the pseudo-spectral method is to discretize the equation, and then transform the discretized equation into a generalized eigenvalue problem, solving for the eigenvalues. Usually, a continuous function is approximately represented by a sequence of its values on collocation points, which  are usually selected as Chebyshev coordinates
\begin{equation}
x_i=\cos\left(\frac iN\pi\right),i=0,...,N .
\end{equation}
The second step is to construct
the Chebyshev basis functions, which can be expressed as:
\begin{equation}
C_j(x)=\prod\limits_{j=0,j\neq i}^N\frac{x-x_j}{x_i-x_j} .
\end{equation}
Finally, the basis function is used to fit the objective function $f(x)$, which can be approximated as follows:
\begin{equation}
\begin{aligned}f(x)&\approx\sum_{j=0}^Nf(x_j)C_j(x) .\end{aligned}
\end{equation}
The Chebyshev basis function can be expressed as:
\begin{equation}
C_j(x)=\frac{2}{Np_j}\sum_{m=0}^N\frac{1}{p_m}T_m(x_j)T_m(x), p_0=p_N=2 , p_j=1.
\end{equation}
In the case of QNMs, due to the particularity of the boundary conditions, it is often convenient to use Eddington coordinates. We make the following substitution

\begin{equation}
r\to\frac{r_h}{u}  \ \ \ \ \ \ \ \ \mathrm{and} \ \ \ \ \ \ \ \  \Psi=e^{i\omega r_*(u)}\psi .
\label{trs}
\end{equation}
In this way, the coordinate $u$ runs from 0 to 1. The wave equation in Eq.(\ref{Sch_like_eq}) under the scalar field perturbation transforms into the following expression:
\begin{equation}\label{sc eq}
\psi''(u)+\left[\frac{f'(u)}{f(u)}+\frac{2ir_h\omega}{u^2f(u)}\right]\psi'(u)-\left[\frac{l(l+1)}{u^2f(u)}+\frac{2ir_h\omega}{u^3f(u)}\right]\psi(u)=0.
\end{equation}
While the wave equation for electromagnetic field perturbations transforms into the following expression:
\begin{equation}\label{ele eq}
\psi''(u)+\left[\frac{f'(u)}{f(u)}+\frac{2ir_h\omega}{u^2f(u)}\right]\psi'(u)-\left[\frac{l(l+1)}{u^2f(u)}+\frac{2ir_h\omega}{u^3f(u)}+\frac{ f'(u)}{u f(u)}\right]\psi(u)=0.
\end{equation}
Eq.(\ref{sc eq}) and Eq.(\ref{ele eq}) satisfy the boundary conditions at the horizon. In order to ensure that there are only outgoing waves at infinity, we take the following transformation 
 for the variable:
\begin{equation}
\psi(u)=e^{\frac{2ir_h\omega}u}u^{2ir_h\omega f'(0)}\delta\psi(u),
\end{equation}
then it can be checked that the boundary condition at infinity is automatically satisfied.  Finally, we obtain the equation for scalar field perturbations and electromagnetic field perturbations, which satisfy the boundary conditions of having only incoming waves at the horizon and only outgoing waves at infinity. This equation is given by:
\begin{equation}\label{B9}
\delta\psi''(u)+\lambda_s(u)\delta\psi'(u)+\mu_s(u)\delta\psi(u)=0.
\end{equation}
 Here, $s=0$ indicates scalar perturbation and $s=1$ indicates electromagnetic perturbation. $\lambda_s$ and $\mu_s$ have the following form:
\begin{equation}
\begin{aligned}
\lambda_{0}(u) &= \frac{f'(u)}{f(u)}+\frac{4ir_h\omega f'(0)}{u}+\frac{2ir_h\omega}{u^2f(u)}-\frac{4ir_h\omega}{u^2} \\
\mu_{0}(u) &= -\frac{l(l+1)}{u^2f(u)}-\frac{2ir_h\omega[1+f(u)(-2+uf^{\prime}(0))+uf^{\prime}(u)(1-uf^{\prime}(0))]}{u^3f(u)} \\
&\quad -\frac{4r_h^2\omega^2(-1+uf'(0))[1+f(u)(-1+uf'(0))]}{u^4f(u)}.
\end{aligned}
\end{equation}

\begin{equation}
\begin{aligned}
\lambda_{1}(u) &= \frac{4 i r_h \omega  f'(0)}{u}+\frac{f'(u)}{f(u)}+\frac{2 i r_h \omega }{u^2 f(u)}-\frac{4 i rh \omega }{u^2} \\
\mu_{1}(u) &= \frac{8 \omega ^2 f'(0) r_h^2}{u^3}-\frac{4 \omega ^2 f'(0) r_h^2}{u^3 f(u)}-\frac{4 \omega ^2 f'(0)^2 r_h^2 + 2 i \omega  f'(0) r_h}{u^2} \\
&\quad -\frac{2 i \omega  r_h f'(u)}{u^2 f(u)}+\frac{2 i \omega  f'(0) r_h f'(u)-f'(u)}{u f(u)}+\frac{4 \omega ^2 r_h^2}{u^4 f(u)}-\frac{2 i \omega  r_h}{u^3 f(u)} \\
&\quad -\frac{l(l + 1)}{u^2 f(u)}-\frac{4 \omega ^2 r_h^2}{u^4}+\frac{4 i \omega  r_h}{u^3}.
\end{aligned}
\end{equation}
Afterward, Eq.(\ref{sc eq}) and  Eq.(\ref{ele eq}) transform into a generalized eigenvalue problem:
\begin{equation}\label{generalized}
(M_0+\omega M_1)\phi=0.
\end{equation}
where $M_i (i=0,1) $ are purely numerical matrices. Then, the frequency of QNMs can be obtained by solving the eigenvalues from Eq.(\ref{generalized}).
\appendix
\bibliographystyle{style1}
\bibliography{qnm}

\end{document}